\newcommand{\bea}{\begin{eqnarray}}
\newcommand{\eea}{\end{eqnarray}}
\newcommand{\be}{\begin{equation}}
\newcommand{\ee}{\end{equation}}
\newcommand{\ben}{\begin{enumerate}}
\newcommand{\een}{\end{enumerate}}
\newcommand{\bi}{\begin{itemize}}
\newcommand{\ei}{\end{itemize}}
\newcommand{\bmi}[1]{\begin{minipage}{#1 cm}}
\newcommand{\emi}{\end{minipage}}

\def\A{\mathcal A}

\newcommand{\rund}[1]{\left(#1\right)}
\newcommand{\vc}[1]{\mbox{\boldmath $#1$}}
\renewcommand{\d}{{\rm d}}
\newcommand{\eck}[1]{\left[ #1 \right]}

\newcommand{\vt}{\vartheta}
\newcommand{\vp}{\varphi}
\newcommand{\abs}[1]{\left| #1 \right|}

\def\elabel#1{\label{eq:#1}}

{\catcode`\@=11
\gdef\SchlangeUnter#1#2{\lower2pt\vbox{\baselineskip 0pt \lineskip0pt
  \ialign{$\m@th#1\hfil##\hfil$\crcr#2\crcr\sim\crcr}}}
}
\def\gtrsim{\mathrel{\mathpalette\SchlangeUnter>}}

\documentclass[onecolumn]{aa}
\usepackage{graphicx}
\usepackage{epsfig}
\begin{document}
   \title{Weak lensing goes bananas: What flexion really measures}

   \author{
   Peter Schneider\inst{1},
   Xinzhong Er\inst{1,2}
   }
   \offprints{P. Schneider}

 \institute{ Argelander-Institut f\"ur Astronomie, Universit\"at Bonn,
              Auf dem H\"ugel 71, D-53121 Bonn, Germany\\
          \email{peter, xer@astro.uni-bonn.de}
  \and
   Max-Plank-Institut f\"ur Radioastronomie,
          Auf dem H\"ugel 69, D-53121 Bonn, Germany}

\titlerunning{What flexion really measures}

   \date{Received ; accepted }

\abstract{In weak gravitational lensing, the image distortion caused
by shear measures the projected tidal gravitational field of the
deflecting mass distribution. To lowest order, the shear is
proportional to the mean image ellipticity. If the image sizes are
not small compared to the scale over which the shear varies,
higher-order distortions occur, called flexion.

For ordinary weak lensing, the observable quantity is not the shear,
but the reduced shear, owing to the mass-sheet degeneracy. Likewise,
the flexion itself is unobservable. Rather, higher-order image distortions
measure the reduced flexion, i.e., derivatives of the reduced
shear. We derive the corresponding lens equation in terms of the
reduced flexion and calculate the resulting relation between
brightness moments of source and image. Assuming an isotropic
distribution of source orientations, estimates for the reduced shear
and flexion are obtained; these are then tested with simulations.  In
particular, the presence of flexion affects the determination of the
reduced shear.  The results of these simulations yield the amount of
bias of the estimators, as a function of the shear and flexion. We
point out and quantify a fundamental limitation of the flexion
formalism, in terms of the product of reduced flexion and source
size. If this product increases above the derived threshold, multiple
images of the source are formed locally, and the formalism breaks
down. Finally, we show how a general (reduced) flexion field can be
decomposed into its four components: two of them are due to a shear
field, carrying an E- and B-mode in general. The other two components
do not correspond to a shear field; they can also be split up into
corresponding E- and B-modes.

   \keywords{cosmology -- gravitational lensing -- large-scale
                structure of the Universe -- galaxies: evolution --
              galaxies: statistics
               }
   }

   \maketitle
%

\section{Introduction}

Weak gravitational lensing provides a powerful tool for studying the
mass distribution of clusters of galaxies as well as the large scale
structure in the Universe (see Mellier 1999; Bartelmann \& Schneider
2001; Refregier 2003; Schneider 2006; Munshi et al.\ 2006 for reviews
on weak lensing).  It has led to constraints on cosmological
parameters, such as those characterizing structure formation and
the mass density of the Universe.

In weak lensing, one employs the fact that the image ellipticity of a
distant source is modified by the tidal gravitational field of the
intervening matter distribution. Based on the assumption that the
orientation of distant sources is random, the ellipticity of each
image yields an unbiased estimate of the line-of-sight integrated
tidal field, usually called shear in lensing. The shear thus carries
information about the properties of the mass
distribution. Formally, the shear is described in terms of a
first-order expansion of the lens equation, i.e., the locally
linearized lens equation. This yields a valid description of the
mapping from the image to the source sphere, as long as the images are
small compared to the length-scale on which the shear varies. However,
this linear approximation breaks down for larger sources, or in
regions of the lens plane where the shear varies rapidly. The most
visible failure of the linearized lens equation is the occurrence of
giant arcs, which in most cases correspond actually to multiple images
of a background source; to model them, the full lens equation needs to
be studied. However, there is an intermediate regime where the
linearized lens equation breaks down, although (locally) no multiple
images are formed -- the arclets regime. Arclets are fairly strongly
distorted images of background sources (Fort et al.\ 1988; Fort \&
Mellier 1994), though they do not correspond to multiple images.

Arclets are the most natural application for flexion. Flexion has been
introduced by Goldberg \& Bacon (2005) and Bacon et al.\ (2006), and
describes the lowest-order deviation of the lens mapping from its
linear expansion. It corresponds to the derivative of the shear; in
combination with a strong shear, this can deform round images into
arclets, giving rise to images which resemble the shape of a
banana. In their original paper, Goldberg \& Bacon (2005) considered
only a single component of flexion which, however, only provides an
incomplete description of shear derivatives. In Bacon et al.  (2006),
the need for a second flexion component was recognized.

In the first part of this paper, we present the general theory of
flexion; in contrast to earlier work, we explicitly consider the
quantities that can be actually observed, by accounting for the
mass-sheet degeneracy (Falco et al.\ 1985; Gorenstein et al.\ 1988).
That is, a change of the surface mass density $\kappa$ of the form
$\kappa \to \lambda\kappa+(1-\lambda)$ leaves the shape of all
observed images invariant. In usual weak lensing, this is accounted
for by recognizing that not the shear $\gamma$ can be obtained from
observations, but only the reduced shear $g=\gamma/(1-\kappa)$
(Schneider \& Seitz 1995). The difference of shear and reduced shear
is typically small, in particular in applications of cosmic shear,
since along most lines-of-sight, the value of $\kappa$ is very much
smaller than unity. In applications of flexion, however, we expect
that the surface mass density no longer is very small; for instance,
arclets occur in the inner parts of clusters where $\kappa\gtrsim
0.1$. Therefore, the difference between shear and reduced shear can no
longer be neglected. Gradients of the shear are not directly
observable; only derivatives of the reduced shear are, and thus we
define the (reduced) flexion in terms of derivatives of $g$.  In
Sect.\ts\ref{Sc:2.1} we briefly recall the irreducible tensor
components which are defined in term of their behavior under rotations
of the coordinate system. It turns out that a complex notation for
these tensor components is very useful.  In Sect.\ts\ref{Sc:2.2} we
expand the lens equation to second order, before deriving the
corresponding lens equation (and relation for the local Jacobian)
which is invariant under mass-sheet transformations. The second-order
term in this lens equation is fully described by our reduced flexion
components $G_1$ and $G_3$.

As is known from usual weak lensing studies, a measured shear is not
necessarily accounted for by an (equivalent) surface mass density.
Since the shear is a two-component quantity, it has one degree of
freedom more than the $\kappa$ field. Therefore, shear fields are
decomposed into E- and B-modes (Crittenden et al.\ 2002; Schneider et
al.\ 2002), where the former are due to a $\kappa$
field, whereas the latter describes the remaining (``curl'') part. A
similar situation occurs in flexion, which has four
components. Therefore, in Sect.\ts\ref{Sc:3} we consider the
decomposition of a general flexion field into contributions due to the
gradient of the shear and those not related to the shear field. The
former one can then be further subdivided into flexion resulting from
an E- and B-mode shear field. We carry out this decomposition for the
flexion as well as for the reduced flexion.

In Sect.\ts\ref{Sc:4} we then define brightness moments of sources and
images and derive the transformation laws between them. This approach
is very similar to the HOLICs approach developed by Okura et al.\
(2006) and later considered by Goldberg \& Leonard (2007), except that
we explicitly write all relations in terms of the reduced shear and
the reduced flexion. Generalizing the usual assumption that the
expectation value of the source ellipticity is zero -- due to the
phase averaging over source orientations -- to the expectation values
of all source shape parameters which are not invariant under
coordinate rotations (as appropriate for a statistically isotropic
Universe), we obtain in Sect.\ts\ref{Sc:5}
estimates for the reduced shear and reduced
flexion in terms of the brightness moments of the images. In
Sect.\ts\ref{Sc:6} we perform a number of numerical experiments to
test the validity of our approach and the accuracy of the estimators
derived. In particular, we point out that there is a fundamental limit
where the theory of flexion has to break down -- the second-order lens
equation is non-linear and will in general have critical curves,
leading to multiple images of the source (or parts of it). If the
source is cut by a caustic, different parts of it will have different
numbers of images, and the assumption of random source orientation
(which underlies all weak lens applications) will break down -- the
caustic introduces a preferred orientation into the source plane. In
appendix B we provide a full classification of the critical curves of
the second-order lens equation and use these results in order to
obtain the maximum source size (for given values of the reduced
flexion) for which the flexion concept still makes sense.
We discuss our results in Sect.\ts\ref{Sc:7}.

\section{\label{Sc:2}Complex lensing notation}
Like in many other instances in weak lensing, flexion is best
described by using complex notation, which we shall briefly introduce
next and which will be used for vectors and tensor components
throughout this paper.

\subsection{\label{Sc:2.1}Irreducible tensor components}
For a two-dimensional vector $\vc
x=(x_1,x_2)$, we define the complex number $x=x_1+{\rm i} x_2$.
Under rotations of the coordinate system by an angle $\vp$, $x$ gets
multiplied by the phase factor ${\rm e}^{-{\rm i}\vp}$. For a tensor
of second rank, whose Cartesian components are $Q_{ij}$, we define
the complex numbers $Q_2=Q_{11}-Q_{22}+{\rm i}(Q_{12}+Q_{21})$ and
$Q_0=Q_{11}+Q_{22}+{\rm i}(Q_{12}-Q_{21})$. A rotation of the
coordinate systems by an angle $\vp$ multiplies $Q_2$ by the phase
factor ${\rm e}^{-2{\rm i}\vp}$, whereas $Q_0$ remains unchanged.
This is most easily seen by considering that the prototype of a
second rank tensor is $Q_{ij}=x_i y_j$, where $\vc x$ and $\vc y$
are vectors; the foregoing statements are then obtained by noting
that the complex numbers $xy$ and $x^* y$ are multiplied by ${\rm
e}^{-2{\rm i}\vp}$ and $1$, respectively, under coordinate rotations.
According to this transformation behavior, we shall loosely speak
about $Q_0$ as a spin-0 quantity, whereas $x$ and $Q_2$ are spin-1
and spin-2 quantities, respectively.

We shall be dealing only with totally symmetric tensors. If $Q_{ij}$
is symmetric, then
\be
Q_2:=Q_{11}-Q_{22}+2{\rm i}Q_{12} \; ; \;\;
Q_0:=Q_{11}+Q_{22}\;.
\elabel{Q02def}
\ee
If $T_{ijk}$ is a symmetric third-rank
tensor, we define its spin-3 and spin-1 components as
\be
T_3:=T_{111}-3T_{122}+{\rm i}\rund{3T_{112}-T_{222}} \; ; \;\;
T_1:=T_{111}+T_{122} +{\rm i}\rund{T_{112}+T_{222}} \;.
\elabel{T13def}
\ee
Furthermore, if $F_{ijkl}$ denotes a symmetric fourth-rank tensor, we
decompose it into its spin-4, spin-2 and spin-0 components, respectively,
\be
F_4:=F_{1111}-6F_{1122}+F_{2222}+4{\rm i}\rund{F_{1112}-F_{1222}}
\, ; \;
F_2:=F_{1111}-F_{2222}+2{\rm i}\rund{F_{1112}+F_{1222}}
\, ; \;
F_0:=F_{1111}+2F_{1122}+F_{2222} \,.
\elabel{F024def}
\ee
Apart from notational simplicity, the complex lensing notation
provides a check for the validity of equations. In a valid equation,
each term has to have the same spin. The product of a spin-$m$ and a
spin-$n$ quantity has spin $m+n$. The complex conjugate of a spin-$n$
quantity has spin $-n$.

\subsection{\label{Sc:2.2}Second-order expansion of the local lens equation}
In weak lensing, the lens equation is linearized locally by writing
the relative source coordinate $\vc \beta$ in terms of the image
position $\vc\theta$ as $\beta_i=\theta_i-\psi_{,ij}\theta_j$, where
$\psi$ is the deflection potential, indices separated by a comma
denote partial derivatives with respect to $\theta_i$, and summation
over repeated indices is implied. Note that the form of this equation
implies that the origin of the lens plane, $\vc\theta=0$, is mapped
onto the origin of the source plane.  The surface mass density
$\kappa$ and the complex shear $\gamma$ at the origin are
given in terms of the deflection potential,
$\kappa=(\psi_{,11}+\psi_{,22})/2$, $\gamma=(\psi_{,11}-\psi_{,22})/2
+{\rm i}\psi_{,12}$, being spin-0 and spin-2 fields, respectively.  In
our complex notation, the locally linearized lens equation reads
\be
\beta=(1-\kappa)\theta-\gamma\theta^*\;.
\elabel{linlens}
\ee
We next generalize this result to a second-order local expansion of
the lens equation, which in Cartesian coordinates reads
$\beta_i=\theta_i-\psi_{,ij}\theta_j-\psi_{,ijk}\theta_j\theta_k$/2.
The third-order derivatives of $\psi$ are related to the gradient of
$\kappa$ and $\gamma$. To write these derivatives also in complex
form, we define the differential operators
\be
\nabla_{\rm c}:={\partial\over \partial \theta_1}+{\rm
i}{\partial\over \partial \theta_2} \; ; \;\;
\nabla_{\rm c}^*:={\partial\over \partial \theta_1}-{\rm
i}{\partial\over \partial \theta_2} \;.
\elabel{gradcdef}
\ee
The differential operator $\nabla_{\rm c}$ turns a spin-$n$ field into
a spin-$(n+1)$ field, whereas $\nabla_{\rm c}^*$ reduces the spin by
one unit. One finds, for example,
\be
\nabla_{\rm c}\kappa={1\over 2}\eck{\psi_{,111}+\psi_{,122}+{\rm
i}\rund{\psi_{,112}+\psi_{,222}}} \; ; \;\;
\nabla_{\rm c}\gamma={1\over 2}\eck{\psi_{,111}-3\psi_{,122}+{\rm
i}\rund{3\psi_{,112}-\psi_{,222}}} \; ; \;\;
\nabla_{\rm c}^*\gamma=\nabla_{\rm c}\kappa\;,
\elabel{kapdiff}
\ee
and we recognize the combinations of third derivatives of $\psi$ which
form the spin-1 and spin-3 combinations defined in (\ref{eq:T13def}).
The final relation in (\ref{eq:kapdiff}) is the relation between first
derivatives of $\kappa$ and $\gamma$ found by Kaiser (1995), here
expressed in compact form. It expresses the fact that the third-order
derivatives of the deflection potential can be summarized in the
spin-3 field ${\cal G}\equiv\nabla_{\rm c}\gamma$ and the spin-1 field
${\cal F}\equiv\nabla_{\rm c}^*\gamma$, where we introduced the usual
notation for the two flexion quantities.  The second-order lens
equation in our complex notation then reads
\be
\beta=(1-\kappa)\theta-\gamma\theta^*
- {1\over 4}{\cal F}^*\,\theta^2
-{1\over 2}{\cal F}\,\theta\theta^*
- {1\over 4}{\cal G}\,(\theta^*)^2 \;.
\elabel{lenseq}
\ee
Since this is no longer a linear equation, a source at $\beta$ may
have more than one image. In fact, up to four images of a source can
be obtained, as can be seen for the special case of $\gamma=0={\cal
F}$ and by placing the source at $\beta=0$. In this case, if we set
${\cal G}=|{\cal G}|{\rm e}^{3{\rm i}\zeta}$, then one solution is
$\theta=0$, and the other three are $\theta=4(1-\kappa)/|{\cal
G}|\,{\rm e}^{{\rm i}\vp}$, with $\vp=\zeta$, $\vp=\zeta+2\pi/3$ and
$\vp=\zeta+4\pi/3$.  Of course, the origin for the occurrence of these
solutions lies in the fact that ${\cal G}$ is a spin-3 quantity.  We
shall later need the Jacobian determinant $\det \A$ of this lens
equation, which is
\bea
\det\A&=&(1-\kappa)^2- \gamma\gamma^*
+\vc\theta\cdot \nabla\eck{(1-\kappa)^2-\gamma\gamma^* } + {\cal
O}(\theta^2) 
\nonumber \\
&=&(1-\kappa)^2- \gamma\gamma^*
-\theta\eck{(1-\kappa){\cal F}^*
+{\gamma^*{\cal F}+\gamma\,{\cal G}^*\over 2}}
-\theta^*\eck{(1-\kappa){\cal F}
+{\gamma^*{\cal G}+\gamma\,{\cal F}^*\over 2}}  + {\cal
O}(\theta^2)\;,
\elabel{detA}
\eea
where the first expression is just the first-order Taylor expansion of
the Jacobian around the origin, and in the second step we made use of
the relation $\vc\theta\cdot\nabla=(\theta\nabla_{\rm
c}^*+\theta^*\nabla_{\rm c})/2$. We point out that (\ref{eq:detA}) is
not the full Jacobian of the lens equation (\ref{eq:lenseq}), but only
its first-order expansion; the full Jacobian contains quadratic terms
in $\theta$. We will return to this important issue further below.

\subsection{\label{Sc:2.3}Accounting for the mass-sheet degeneracy}
The observables of a gravitational lens system are unchanged if the
surface mass density $\kappa$ is transformed as $\kappa(\vc\theta)
\to \kappa'(\vc\theta)=\lambda\kappa(\vc\theta) +(1-\lambda)$
(Gorenstein et al.\ 1988). In the case of weak lensing, the shape of
images is unchanged under this transformation (Schneider \& Seitz
1995). Because of this mass-sheet degeneracy, not the shear is an
observable in weak lensing, but only the reduced shear
$g=\gamma/(1-\kappa)$. In fact, since we expect that the most
promising applications of flexion will come from situations where
$\kappa$ is not much smaller than unity, the distinction between shear
and reduced shear is likely to be more important for flexion than for
the usual weak lensing applications.  Hence, at best we can expect
from higher-order shape measurements to obtain an estimate for the
reduced shear and its derivatives. For this reason, we shall rewrite
the foregoing expressions in terms of the reduced shear.

The mass-sheet transformation is equivalent to an isotropic scaling of
the source plane coordinates. Hence, we divide (\ref{eq:lenseq}) by
$(1-\kappa)$ to obtain
\be
\hat\beta\equiv{\beta\over (1-\kappa)}=\theta-g\theta^*
- \Psi_1^*\,\theta^2
-2\Psi_1  \,\theta\theta^*
- \Psi_3 \,(\theta^*)^2 \; {\rm with} \;\;
\Psi_1={1\over 4}{{\cal F}\over(1-\kappa)}
\; ; \; \;
\Psi_3={1\over 4}{{\cal G}\over (1-\kappa)}\;.
\elabel{lenseqred}
\ee
We will now express the coefficients in the lens equation
(\ref{eq:lenseqred}) in terms of the derivatives of the reduced shear,
\be
G_1\equiv \nabla_{\rm c}^* g={{\cal F}+g{\cal F}^*\over (1-\kappa)}
\; ;\quad G_3\equiv \nabla_{\rm c}
g={{\cal G}+g{\cal F}\over (1-\kappa)} \;.
\elabel{G13def}
\ee
The expression for ${\cal F} / (1-\kappa)$ in terms of the
reduced shear and its derivatives has been derived by Kaiser (1995);
in our notation it reads
\be
{{\cal F}\over(1-\kappa)}\equiv -\nabla_{\rm c} \ln(1-\kappa)
={G_1-g G_1^* \over 1-gg^*}
\;
\Rightarrow \;
\Psi_1={G_1-g G_1^*\over 4\rund{1-gg^*}}
 \;.
\elabel{kapredu}
\ee
The expression for the derivative of $\gamma$ in terms of the reduced
shear can be easily obtained from differentiating the definition
$\gamma=(1-\kappa)g$,
\be
{\nabla_{\rm c}\gamma\over (1-\kappa)}={{\cal G}\over (1-\kappa)}
=G_3-g {\nabla_{\rm c}\kappa\over(1-\kappa)}
=G_3 -
{g\rund{G_1-g G_1^*}\over 1-g g^*} \;
\Rightarrow \;\;
\Psi_3={G_3\over 4} -
{g\rund{G_1-g G_1^*}\over 4\rund{1-g g^*}} \; .
\elabel{gamredu}
\ee
The derivatives $G_{1,3}$ of the reduced shear are those quantities we
can hope to observe; to distinguish them from ${\cal F}$ and ${\cal
G}$, one might call $G_{1,3}$ the {\it reduced flexion.}

The Jacobian determinant $\det\hat\A$ of the mapping between the image
position $\theta$ and the rescaled source position $\hat \beta$ then
becomes
\be
\det\hat\A={\det\A\over (1-\kappa)^2}
=1-g g^* -\eta^*\theta-\eta\theta^* \; ,\; {\rm where}
\;\;\;
\eta=\nabla_{\rm c}^*g - {g \rund{\nabla_{\rm c}^*g}^*\over 2}
+{g^* \nabla_{\rm c} g \over 2}= G_1 -{g G_1^* \over 2} + {g^* G_3 \over 2}
\elabel{detAhat}
\ee
is a spin-1 quantity. Again, (\ref{eq:detAhat}) is valid only to
linear order in $\theta$. Note that a similar equation for the
determinant was obtained in Okura et al.\ (2007; their eq.\ts A1), but
they consider only the case of $|g|\ll 1$; this has also consequences
for the relations between source and image brightness moments, to be
derived further below. 

\section{\label{Sc:3}Compatibility relations}
Flexion has a total of four components, namely the real and imaginary
parts of ${\cal F}$ and ${\cal G}$. A measurement of flexion will thus
yield four components, and we might ask whether these components are
independent. We recall a similar situation in shear measurements. The
shear has two components; on the other hand, the shear is defined as
second partial derivatives of the deflection potential, which is a
single scalar field. Therefore, the two shear components cannot be
mutually independent if they are due to a gravitational lensing
signal. Of course, the measured shear is not guaranteed to satisfy the
condition that the two shear components can be derived from a single
scalar deflection potential, since observational noise or intrinsic
alignments of galaxies may affect the measured shear field. Therefore,
one has introduced the notion of E- and B-modes in shear measurements
(Crittenden et al.\ 2002). The E-mode shear is the one that can be
written in terms of a deflection potential, whereas the B-mode shear
cannot. 


Formally, the E- and B-mode decomposition can be written in terms of a
complex deflection potential $\psi(\theta)=\psi^{\rm
E}(\theta)+{\rm i}\psi^{\rm B}(\theta)$ and a complex surface mass
density $\kappa=\kappa^{\rm E} + {\rm i}\kappa^{\rm B}$ (Schneider et
al.\ 2002). Each component of $\psi$ satisfies its own Poisson
equation, $\nabla^2\psi^{\rm E}=2\kappa^{\rm E}$, $\nabla^2\psi^{\rm
B} =2\kappa^{\rm B}$. Making use of this decomposition, the shear
becomes
\be
\gamma=\gamma_1+{\rm i}\gamma_2=\rund{\psi_{,11}-\psi_{,22}}/2+{\rm
  i}\psi_{,12}
=\eck{{1\over 2}\rund{\psi^{\rm E}_{,11}-\psi^{\rm E}_{,22}}
-\psi^{\rm B}_{,12}} +{\rm i}\eck{\psi^{\rm E}_{,12}
+{1\over 2}\rund{\psi^{\rm B}_{,11}-\psi^{\rm B}_{,22}}} \;.
\elabel{gammaEB}
\ee
The distinction between E- and B-mode shear can be obtained by
considering second partial derivatives of the shear components.
Taking the derivative of (\ref{eq:gammaEB}), one obtains
\be
{\cal F}=\nabla_{\rm c}^*\gamma
=(1/2)\rund{\psi^{\rm E}_{,111}+\psi^{\rm E}_{,122}-
\psi^{\rm B}_{,112}-\psi^{\rm B}_{,222} }
+({\rm i}/2)\rund{\psi^{\rm E}_{,112}+\psi^{\rm E}_{,222}+
\psi^{\rm B}_{,111}+\psi^{\rm B}_{,122} }
=\kappa^{\rm E}_{,1}-\kappa^{\rm B}_{,2}
+{\rm i}\rund{\kappa^{\rm E}_{,2}+\kappa^{\rm B}_{,1}} \;,
\ee
which can be expressed in more compact form as
\be
{\cal F}=\nabla_{\rm c}\rund{\kappa^{\rm E}+{\rm i}\kappa^{\rm B}}
=\nabla_{\rm c}\kappa\;.
\ee
A further derivative yields for the components
\be
{\cal F}_{1,1}=\kappa^{\rm E}_{,11}-\kappa^{\rm B}_{,12}\;;\quad
{\cal F}_{1,2}=\kappa^{\rm E}_{,12}-\kappa^{\rm B}_{,22}\;;\quad
{\cal F}_{2,1}=\kappa^{\rm E}_{,12}+\kappa^{\rm B}_{,11}\;;\quad
{\cal F}_{2,2}=\kappa^{\rm E}_{,22}+\kappa^{\rm B}_{,12}\;.
\ee
However, it is easier to consider directly the complex derivative of
${\cal F}$, from which we obtain
\be
\nabla^*_{\rm c}{\cal F}=\nabla^*_{\rm
c}\nabla^*_{\rm c}\gamma=
\nabla^2\rund{\kappa^{\rm E}+{\rm i}\kappa^{\rm
B}}={\cal F}_{1,1}+{\cal F}_{2,2}
+{\rm i}\rund{{\cal F}_{2,1}-{\cal F}_{1,2}} \;.
\ee
Thus, if the shear field is a pure E-mode field, $\nabla^*_{\rm
c}\nabla^*_{\rm c}\gamma$ is real. An imaginary part of $\nabla^*_{\rm
c}\nabla^*_{\rm c}\gamma$ is due to a B-mode field. This then yields the
local distinction between E- and B-mode shear.

Since the flexion has four components, whereas the lens can be
described by a single scalar field, we expect that there are three
constraint relations a flexion field has to satisfy if it is due to a
lensing potential. In fact, even if we leave the shear field arbitrary
(that is, even if we allow it to be composed of E- and B-modes), then
we expect two constraint equations, since the flexion field has two
components more than the shear field. These constraint equations are
easy to obtain. First, if the flexion field is due to a shear field,
then we have
\be
\nabla_{\rm c}\nabla^*_{\rm c}\gamma=\nabla^*_{\rm c}\nabla_{\rm
c}\gamma
\quad{\rightarrow}\quad {\cal H}:=
\nabla_{\rm c}{\cal F}-\nabla^*_{\rm c}{\cal G}=0\;,
\elabel{GFcompat}
\ee
where we defined the spin-2 quantity ${\cal H}$. It may describe
contributions to the
flexion which are not caused by a shear field, such as due
to noise, intrinsic source alignments or higher-order terms (such as
lens-lens coupling) in the propagation equation for light bundles. As
a spin-2 field, a non-zero ${\cal H}$ can be decomposed into its E-
and B-modes.\footnote{Let $H(\theta)$ be any spin-2 field, and
denote by $H^{\rm E}$ and $H^{\rm B}$ the E- and B-mode components of
$H$. They can be obtained from $H$ most easily in Fourier space,
namely
\[
\hat H^{\rm E}(\ell)={1\over 2}\eck{\hat H(\ell)+\hat
H^*(-\ell)\, {\rm e}^{4{\rm i}\beta}} \; ;\quad
\hat H^{\rm B}(\ell)={1\over 2}\eck{\hat H(\ell)-\hat
H^*(-\ell)\, {\rm e}^{4{\rm i}\beta}} \; ,
\]
as can be best seen by taking the shear field as a prototypical spin-2
field; here, $\beta$ is the phase of the complex wave number $\ell$.} 
If ${\cal H}\equiv 0$, then the spin-3 flexion ${\cal G}$
is completely determined by the spin-1 flexion ${\cal F}$ up to an
additive constant, as can be best seen in Fourier space, for which
(\ref{eq:GFcompat}) yields $\hat{\cal G}(\ell)=-{\rm
i}\hat\gamma(\ell)\,\ell =(\ell/\ell^*)\hat{\cal F}(\ell)$.  Second,
if the flexion field is solely caused by a gravitational lens effect,
i.e., by a pure E-mode shear field, then $\nabla^*_{\rm c}{\cal F}$ is
real, i.e.,
\be
{\cal F}_{\rm i}:=\nabla^*_{\rm c}{\cal F}-\nabla_{\rm c}{\cal F}^*=0\;.
\ee
Thus, flexion from a pure E-mode shear field is characterized by the
three constraint equations ${\cal H}\equiv 0$ and ${\cal F}_{\rm
i}\equiv0$, where the former is a two-component equation.

Turning now to the reduced flexion, the compatibility equations can be
obtained as follows. First, if the flexion is due to a shear field, we
have
\be
H:=\nabla_{\rm c}G_1-\nabla^*_{\rm c}G_3=0\;,
\ee
as follows from the definition (\ref{eq:G13def}) of the two flexion
components in terms of the reduced shear. Again, if this equation is
satisfied, $G_3$ is completely determined by $G_1$, up to an additive
constant. Second, if the flexion is caused by a pure E-mode shear,
i.e., if the shear is due to a real surface mass density, then we
employ the
quantity $\ln (1-\kappa)$, which is real and invariant under
mass-sheet transformations, up to an additive constant. Therefore,
$K_2\equiv-\nabla^*_{\rm c}\nabla_{\rm c}\ln (1-\kappa)$ must be
real. We find:
\be
K_2=\nabla_{\rm c}^*\rund{\nabla_{\rm c}\kappa\over 1-\kappa}
= \nabla_{\rm c}^*\eck{{1\over 1-gg^*}\rund{G_1-G_1^*g}}
={\eck{\nabla_{\rm c}^*G_1-g\rund{\nabla_{\rm c}G_1}^*}\over 1-gg^*}
+{\rund{G_1^2 g^*+g G_1 G_3^* - G_1 G_1^* -g^2G_1^*
G_3^*}\over (1-gg^*)^2} \;,
\elabel{22}
\ee
so that a flexion coming from an E-mode shear field satisfies
$K_2=K_2^*$.

We point out that the foregoing relation suggests a natural way to use
flexion for finite-field mass reconstructions in weak lensing. Seitz
\& Schneider (2001) formulated the finite-field mass reconstruction
from measured reduced shear in terms of a von Neumann boundary value
problem for $K=-\ln (1-\kappa)$, whose solution determines $K$ up to
an additive constant. The `source' for $\nabla^2 K$ was determined by
the reduced shear and its derivatives, and is given by
(\ref{eq:22}). In Seitz \& Schneider (2001), the derivatives of the
reduced shear were obtained by finite differencing of $g$. If flexion
is measured, one can replace the `source' for $\nabla^2 K$ by a
weighted sum of the differentiated reduced shear field and the
combination $(K_2+K_2^*)/2$ of the flexion field, with
the weights chosen according to the estimated noise properties of both
contributions.

\section{\label{Sc:4}Brightness moments of source and image}
We consider an image of a source, and denote the brightness
distribution of the source by $I^{\rm s}(\beta)$. Since surface
brightness is conserved by lensing, the brightness distribution of the
image is $I(\theta)=I^{\rm s}(\beta(\theta))$.
Since the scaling of the source plane is
unobservable, we shall only work in the following in terms of the
scaled source plane coordinates, and therefore drop the hat on
$\beta$, as well as on $\A$.

We define the origin of the image (or lens) plane as the
center-of-light of the image under consideration, i.e. we require
\be
\int\d^2\theta\;\theta\,I(\theta) =0\;.
\elabel{center}
\ee
Let $F(\beta)$ be a function of the source coordinate; we define the
operator ${\rm Mom}[F(\beta)]$ as
\be
{\rm Mom}[F(\beta)]=\int \d^2\beta\; F(\beta)\,I^{\rm s}(\beta)
=\int \d^2\theta\;{\det\A}(\theta)\,F(\beta(\theta))\,I(\theta)
\approx\int \d^2\theta\;\rund{1-g g^* -\eta^*\theta-\eta\theta^*}
\,F(\beta(\theta))\,I(\theta) \;,
\elabel{Momdef}
\ee
where here and in the following, we use the linear approximation for
$\det \A$. In particular, setting $F=1$, one finds that
\be
{\rm Mom}[1]\equiv S_0 = \int \d^2\beta\; I^{\rm s}(\beta)
=  \int \d^2\theta\;\rund{1-g g^* -\eta^*\theta-\eta\theta^*}
\,I(\theta) =  \rund{1-g g^*}\,S =\det\A_0 \, S\;,
\elabel{Mom1}
\ee
since first-order moments of the light distribution in the lens plane
vanish, due to our choice (\ref{eq:center}) of the coordinate
system. Here, $S$ is the flux of the lensed image, so that
$S=S_0/\det\A_0$, as usual, where $\det\A_0$ is the Jacobian at the
origin $\theta=0$.

\subsection{Centroid shift}
The origin of the coordinates in the source plane is the image of the
origin in the lens plane as mapped with the lens equation. In
particular, this does not coincide with the center-of-light of the
source, which is given by $\bar\beta\equiv {\rm Mom}[\beta]/S_0$, or
\be
{\bar\beta}={1\over S_0}\int \d^2\beta\; \beta\,I^{\rm s}(\beta)
={1\over S\rund{1-gg^*}}
\int \d^2\theta\;\rund{1-g g^* -\eta^*\theta-\eta\theta^*}
\eck{\theta-g\theta^*
- \Psi_1^*\,\theta^2
-2\Psi_1  \,\theta\theta^*
- \Psi_3 \,(\theta^*)^2 }
\,I(\theta) \;.
\elabel{betabar}
\ee
Expanding the integrand, we note that terms linear in $\theta$ vanish,
due to (\ref{eq:center}).
Defining the second-order brightness moments of the image in the form
\be
Q_2\equiv {1\over S}\int\d^2\theta\;\theta^2\,I(\theta) \; ; \;\;
Q_0\equiv {1\over S}\int\d^2\theta\;\theta\,\theta^*\,I(\theta) \; ,
\elabel{Qdef}
\ee
we obtain for the source centroid shift
\be
\bar\beta={3 G_1 g^*-5 G_1^*-2 g G_3^*\over 4(1-g g^*)} Q_2
+{4 g G_1^* + g^2 G_3^*-G_3 g^* -G_1 (3+g g^*)\over 2(1-g g^*)} Q_0
+{5 g G_1 -3 g^2 G_1^* -(1-3 g g^*) G_3\over 4(1-g g^*)} Q_2^* \;.
\elabel{betabarQ}
\ee
We now write these equations in a more compact form; for this, we
define the matrix $\tens{G}$ by $\tens{G}^{\rm
T}=(G_3^*,G_1^*,G_1,G_3)$, where the `T' denotes the transpose of the
matrix. Then,
\be
\bar\beta=\tens{B} \tens{G}\;,
\elabel{betabarshort}
\ee
where the coefficients of $\tens{B}=(b_1,b_2,b_3,b_4)$ are given by
\bea
b_1&=&{ g^2 Q_0-g Q_2\over 2(1-g g^*)} \; ;\;\;
b_2={8g Q_0-5 Q_2-3 g^2 Q_2^*\over 4(1-g g^*)} \; ;\nonumber \\
b_3&=&{3 g^* Q_2-2(3+g g^*)Q_0+5 g Q_2^*\over 4(1-g g^*)} \; ;\;\;
b_4={(3 g g^* -1)Q_2^*-2 g^* Q_0 \over 4(1-g g^*)} \; .
\elabel{bcoeff}
\eea
The centroid shift in the source plane is thus given by the product of
the derivatives of the reduced shear (expressed by $G_1$ and $G_3$)
and the area of the image, which is proportional to $Q_0$ and
$Q_2$. Of course, since the reduced shear and its derivatives are not
directly observable, the centroid shift in unobservable as well.  To
get an order-of-magnitude estimate of $\bar\beta$, we assume that the
source has a linear angular size $\Theta_{\rm s}$, consider the
reduced shear to be of order unity, and let $\Theta_{\rm c}$ be the
angular scale on which the reduced shear varies. Then,
\be
G_n={\cal O}\rund{1\over \Theta_{\rm c}} \; ;\;\;
Q_n={\cal O}\rund{\Theta_{\rm s}^2}\; \; \Rightarrow \;\;
\bar\beta={\cal O}\rund{\Theta_{\rm s}^2\over \Theta_{\rm c}} \;.
\elabel{betabarest}
\ee

\subsection{Transformation of second-order brightness moments}
Next we consider the second-order brightness moments of the source,
defined as $Q_2^{\rm s}={\rm Mom}[(\beta-\bar\beta)^2]/S_0
={\rm Mom}[\beta^2]/S_0-\bar\beta^2$ and $Q_0^{\rm s}={\rm Mom}[(\beta-\bar\beta)(\beta-\bar\beta)^*]/S_0={\rm Mom}[\beta\beta^*]/S_0-\bar\beta\bar\beta^*$.
By defining the third-order
brightness moments of the image through
\be
T_3\equiv {1\over S}\int\d^2\theta\;\theta^3\,I(\theta) \; ; \;\;
T_1\equiv {1\over S}\int\d^2\theta\;\theta^2\,\theta^*\,I(\theta) \; ,
\elabel{T3T1def}
\ee
we obtain
\bea
Q_2^{\rm s}&=&Q_2 - 2g Q_0 + g^2 Q_2^*
+{2 g^* G_1-3 G_1^*-g G_3^*\over 2(1-g g^*)} T_3
+{8 g G_1^*-(4+3 g g^*)G_1-g^* G_3+2 g^2 G_3^*\over2(1-g g^*)} T_1
\nonumber \\
&+&{(7+g g^*)g G_1-7g^2 G_1^* +(3 g g^*-1) G_3-g^3 G_3^*
\over2(1-g g^*)} T_1^*
+{(1-2 g g^*)g G_3-3 g^2 G_1 +2 g^3 G_1^*\over2(1-g g^*)} T_3^*
-\bar\beta^2\;,
\elabel{Q2s}
\eea
\bea
Q_0^{\rm s}&=& -g^* Q_2+(1+g {g^*})Q_0-g Q_2^*
+{6 g^* G_1^* + (3 g g^*-1) G_3^* - 4 {g^*}^2 G_1\over 4(1-g g^*)} T_3
\nonumber \\
&+&{2 {g^*}^2 G_3+(11+3 g g^*)g^* G_1-(7 +9g g^*)G_1^*-(1+3 g g^*)g G_3^*
\over 4(1-g g^*)} T_1
\nonumber \\
&+&
{2 g^2 G_3^* + (11+ 3 g g^*) g G_1^* - (1+3 g g^*)g^* G_3  -(7+9 g
g^*)G_1\over 4(1-g g^*)} T_1^*
+{6 g G_1-4 g^2 G_1^* - (1- 3 g g^*)G_3\over 4(1-g g^*)} T_3^*
-\bar\beta\bar\beta^*
\elabel{Q0s}
\eea
Note that $Q_0^{\rm s}$ is real. In a more compact notation,
(\ref{eq:Q2s}) reads 
\be
Q_2^{\rm s}=Q_2 - 2g Q_0 + g^2 Q_2^*
+A \tens{G}-{\bar\beta}^2\;,
\elabel{Q2short}
\ee
where the matrix $\tens{A}=(a_1,a_2,a_3,a_4)$ has coefficients
\bea
a_1&=&{-g^3 T_1^*+2 g^2 T_1 -g T_3
\over 2(1-gg^*)}\; ; \;\;
a_3={-3 g^2 T_3^*+g(7+ g g^*) T_1^* -(4+3 g g^*)T_1+2 g^*T_3
\over 2(1-gg^*)}\; ;  \nonumber \\
a_2 &=& {2 g^3 T_3^* -7 g^2 T_1^* + 8 g T_1 -3 T_3
\over 2(1-gg^*)}\; ; \;\;
a_4={g(1-2 g g^*)T_3^*-(1- 3 g g^*)T_1^* - g^* T_1
\over 2(1-gg^*)}\; .
\elabel{acoeff}
\eea

\subsection{Transformation of third-order brightness moments}
We now define the third-order brightness moments of the source,
separated into a spin-3 and a spin-1 component,
\be
T_3^{\rm s}={ {\rm Mom}[\rund{\beta-\bar\beta}^3]\over S_0}
={ {\rm Mom}[\beta^3]\over S_0}-3\bar\beta\,{ {\rm Mom}[\beta^2]\over
S_0} +3\bar\beta^2\,{ {\rm Mom}[\beta]\over
S_0} -\bar\beta^3
={ {\rm Mom}[\beta^3]\over S_0}-3\bar\beta Q_2^{\rm s}
- \bar\beta^3 \;,
\elabel{T3def}
\ee
where we used that ${\rm Mom}[\beta^2]/S_0=Q_2^{\rm
s}+\bar\beta^2$
and ${\rm Mom}[\beta\beta^*]/S_0=Q_0^{\rm s}+\bar\beta \bar\beta^*$.
Similarly, we obtain
\be
T_1^{\rm s}={ {\rm Mom}[\rund{\beta-\bar\beta}^2
 (\beta^*-\bar\beta^*)]\over S_0}
={ {\rm Mom}[\beta^2\beta^*]\over S_0}-2Q_0^{\rm s}\bar\beta
-Q_2^{\rm s}\bar\beta^*-\bar\beta^2\bar\beta^* \;.
\elabel{T1def}
\ee
Defining the fourth-order brightness moments of the image by
\be
F_0={1\over S}\int\d^2\theta\;(\theta \theta^*)^2 I(\theta)\; ;\;\;
F_2={1\over S}\int\d^2\theta\; \theta^3 \theta^*\, I(\theta)\; ;\;\;
F_4={1\over S}\int\d^2\theta\; \theta^4\,  I(\theta)\; ,
\elabel{Fdef}
\ee
where $F_n$ is a spin-$n$ quantity, we obtain for the third-order
moments of the source:
\be
{\cal T}^{\rm s} = \tau + \tens{C}\,\tens{G}\,+{\cal O}(\bar\beta ^3)\;,
\elabel{Tsource}
\ee
where the matrix ${\cal T}^{\rm s}$ is defined by its transpose
${\cal T}^{\rm s, T}=\rund{ T_3^{s*},T_1^{s*}, T_1^{s}, T_3^{s}}$.
The elements of $\tau$ are
\be
\tau_1=T_3^*-3 g^* T_1^*+3 g^{*2} T_1- g^{*3} T_3\; ;\;\;
\tau_2=-g T_3^* +(1+2 g g^*)T_1^* - g^*(2 +g g^*)T_1 +g^{*2} T_3\; ;
\;\;
\tau_3=\tau_2^*\; ;\;\; \tau_4=\tau_1^* \;,
\elabel{taus}
\ee
where the last two relations are obvious. The $4\times 4$ matrix
$\tens{C}$ is given explicitly in Appendix\,\ref{Sc:A}; each of its
elements consists of a sum of terms proportional to fourth-order
brightness moments, $F_n$, and terms proportional to squares of
second-order brightness moments. Okura et al.\ (2007) and Goldberg \&
Leonard (2007) have derived expressions similar to (\ref{eq:Tsource}),
though using a number of simplifying assumptions (such as $|g|\ll 1$)
and (in the latter paper), not considering the reduced flexion.

We will now consider the order-of-magnitudes of the various terms
appearing in (\ref{eq:Q2short}) and (\ref{eq:Tsource}). Assuming that
the third-order moments of the sources are small, then the third-order
moments of the image are given by the product of $\tens{C}$ and $\tens{G}$.
With $\tens{G}={\cal O}\rund{1/\Theta_{\rm c}}$ and $\tens{C}={\cal O}
\rund{\Theta_{\rm s}^4}$, we find that
$T={\cal O} \rund{\Theta_{\rm s}^4/\Theta_{\rm c}}= {\cal O}
\rund{\Theta_{\rm s}^3}\, \rund{\Theta_{\rm s}/\Theta_{\rm c}}$. To
get an estimate of the size of the various terms in
(\ref{eq:Q2short}), we note that the first three terms on the
right-hand side (those proportional to the $Q_n$) are of order ${\cal
O}\rund{\Theta_{\rm s}^2}$, whereas $\tens{A}\tens{G}={\cal
O}\rund{\Theta_{\rm s}^4/\Theta_{\rm c}}
\,{\cal O}\rund{1/\Theta_{\rm c}}$ and $\bar\beta^2={\cal
O}\rund{\Theta_{\rm s}^4/\Theta_{\rm c}^2} $. Hence, the last two
terms are of equal magnitude in general, each of them being smaller
than the first three terms by a factor $\rund{\Theta_{\rm
s}/\Theta_{\rm c}}^2$. Only if the source is of the same order as the
scale over which the reduced shear varies do the last two terms in
(\ref{eq:Q2short}) contribute. In (\ref{eq:Tsource}), we have
neglected the terms $\bar\beta^3$, since they are two powers of
$\rund{\Theta_{\rm s}/\Theta_{\rm c}}$ smaller than the terms written
down.

\section{\label{Sc:5}Shear and flexion estimates}
\subsection{Estimate of the reduced shear}
We see that (\ref{eq:Tsource}) is a linear equation for $\tens{G}$,
which can thus be solved,
\be
\tens{G}=\tens{C}^{-1}\rund{ {\cal T}^{\rm s}-\tau }\; .
\elabel{g13}
\ee
Inserting this into (\ref{eq:Q2short}) then yields
\be
Q_2^{\rm s}=Q_2 - 2g Q_0 + g^2 Q_2^*
+\tens{A}\, \tens{C}^{-1}\rund{ {\cal T}^{\rm s}-\tau }
-\rund{\tens{B}\tens{G}}^2\; . 
\ee
We are thus left with a single complex equation for $g$, which
contains the observable brightness moments of the image, as well as
the unobservable brightness moments of the source. This equation can
be used to estimate the reduced shear if we make assumptions concerning
the properties of the source brightness moments. We assume that the
sources are oriented randomly, which implies that all quantities with
spin unequal zero have a vanishing expectation value. That is, we set
$Q_2^{\rm s}=0$, ${\cal T}^{\rm s}=0$, to arrive at
\be
Q_2 - 2g Q_0 + g^2 Q_2^*
= \tens{A}\, \tens{C}^{-1}\tau + \rund{\tens{B}\tens{C}^{-1}\tau}^2 =:
Y(g)\; , 
\elabel{gequ}
\ee
where we have indicated that the right-hand side depends on the
reduced shear (in fact it does so in a very complex manner). However,
since we have argued above that the terms on the left-hand are much
larger than those on the right-hand side, an iterative solution of
this equation is suggested. Assume the right-hand side is given, then
we get the solutions
\be
g={\chi\over |\chi|^2}\rund{1\pm\sqrt{1-|\chi|^2+{Y\chi^*\over Q_0}}}
\;, \;\; {\rm where}\;\; \chi={Q_2\over Q_0}
\ee
is the complex ellipticity of the image. Obviously, there are two
solutions $g$ for a given value of $Y$. This situation is similar to
that of `ordinary' weak lensing, where this ambiguity also occurs: as
shown by Schneider and Seitz (1995), from shape measurements of
background galaxies, and cannot distinguish locally between an estimate
$g$ and $1/g^*=g/|g|^2$. The same occurs here; we therefore assume
that we pick one of the two solutions, say the one corresponding to
the `$-$' sign; this then yields for small shear $g\approx \chi/2$. It
should be stressed that flexion impacts the determination of shear
from the second-order brightness moments, due to its impact on
higher-order brightness moments; hence, in general the
determination of shear and flexion are coupled.

We start the iteration by setting $Y_0=0$. This yields a first-order
solution for the estimate of $g$,
\be
g_0={\chi\over |\chi|^2}\rund{1-\sqrt{1-|\chi|^2}}\;.
\ee
We then use the iteration equations
\be
Y_n=Y(g_{n-1}) \; ; \;\; g_n={\chi\over
|\chi|^2}\rund{1-\sqrt{1-|\chi|^2+{Y_n\chi^*\over Q_0}}} \;.
\ee
This procedure converges quickly to one of the two solutions
$(g,G_1,G_3)$; the other solution is obtained by taking the `+' sign
in the above equations.

Of course, our approach of setting $Q_2^{\rm s}=0$ yields a biased
estimator for $g$; this is true even in the absence of flexion (e.g.,
Schneider \& Seitz 1995). The reason is that, although the expectation
value of $Q_2^{\rm s}$ vanishes, the resulting estimator for $g$ is a
non-linear function of $\chi^{\rm s}=Q_2^{\rm s}/Q_0^{\rm s}$ and thus
biased. The bias depends on the ellipticity distribution of the
sources. It should be stressed, however, that a modified definition of
image ellipticity exist such that its expectation value is an unbiased
estimate of the reduced shear (Seitz \& Schneider 1997).

\subsection{Estimates for the reduced flexion}
The flexion estimator is given by (\ref{eq:g13}). Since the matrix
$\tens{C}$ contains many terms, this is a fairly complicated equation
in general. A simpler estimate is obtained if we assume that the
reduced shear is small, $|g|\ll 1$, in which case the matrix
$\tens{C}$ simplifies considerably -- see Appendix. Furthermore, if we
assume that the brightness moments of spin $\ne 0$ are much smaller
than the corresponding ones with spin 0, then we find the simple
relations
\be
T_1^{\rm s} \approx  T_1-{9 F_0-12 Q_0^2 \over 4}G_1 \; ;\quad
T_3^{\rm s} \approx  T_3- {3 F_0 \over 4} G_3 \;.
\ee
If we then set the $T_n^{\rm s}=0$, as would be true for the
expectation value, then we obtain as estimates for the reduced flexion
\be
G_1 \approx {4 \over 9F_0-12Q_0^2}T_1\; ; \quad
G_3 \approx {4\over 3 F_0}T_3.
\elabel{eg13}
\ee
Thus, the flexion is then given by the third-order brightness moments
of the image, divided by a quantity that just depends on the size of
the image. Similar relations to (\ref{eq:eg13}) have been given in
Goldberg \& Leonard (2007), whereas Okura et al.\ (2007) obtain a
different expression for $G_1$. We will check the accuracy of
(\ref{eq:eg13}) in Sect.\ts\ref{Sc:6} below.

A more accurate estimate is obtained if we consider the reduced shear
as well as the ratios of non-zero spin brightness moments to zero spin
moments (such as $|Q_2/Q_0|$ or $|F_{2,4}/F_0|$) to be of order
$\delta$, and then expand the flexion to first order in the (small)
parameter $\delta$ to obtain
\bea
G_1\!\!&=&\!\!{4T_1\over  9F_0-12Q_0^2}
\!+\!{4\eck{2 F_2^*+3 F_0g^* -2 Q_0(2 g^*Q_0+Q_2^*)}\over
9 F_0(4Q_0^2-3F_0)} T_3
\!+\!{4\eck{3 F_0 g-8 F_2 -4 Q_0(g Q_0 -4 Q_2)} \over
9(3 F_0-4 Q_0^2)^2} T_1^*
\!+\!{4 F_4 T_3^*\over 9 F_0(4 Q_0^2-3 F_0)} \;,\nonumber \\
G_3\!\!&=&\!\!{4 T_3\over 3 F_0}+{8(5 F_2-9 Q_0Q_2)T_1
\over9 F_0(4 Q_0^2-3 F_0)} +{28 F_4 T_1^*\over 9 F_0(4 Q_0^2-3 F_0)} \;.
\elabel{eg13long}
\eea

\section{\label{Sc:6}Numerical tests of flexion estimators}
In this section we describe some simulations that we have performed in
order to test the behavior of the estimators given in the previous
section.

\subsection{Description of the simulations}
We model the sources as elliptical Gaussians, truncated at three times
the scale `radius' $\Theta_{\rm s}$ chosen such that the area of a
source is independent of its ellipticity.
The ellipticity of the sources
follows a Gaussian distribution, with a dispersion of $\chi^{\rm s}$
of $R=0.4$ (i.e., we use the same ellipticity distribution as in
Schneider \& Seitz 1995). However, for reasons explained in the next
subsection, we truncate the intrinsic ellipticity distribution at
$|\chi^{\rm s}|\le 0.9$. For each source, we map a grid of pixels from
the lens plane to the source plane using the lens equation to obtain
the brightness distribution in the lens plane. From this distribution,
the brightness moments of the image are measured. A shift in the lens
plane coordinates is applied as to satisfy (\ref{eq:center}). We then
apply the shear and flexion estimators described above to the resulting
brightness moments $Q_n$, $T_n$ and $F_n$. The shear and flexion
estimates are then averaged over the Gaussian ellipticity distribution
of the sources, in particular over their random orientation.

It should be noted that flexion is a dimensional quantity $\propto
\Theta_{\rm c}^{-1}$. As can be checked explicitly from
Sect.\ts\ref{Sc:4}, the way flexion appears in the equations is always
with one order higher in the source (or image) size than the other
terms in the equations. As an example, we consider (\ref{eq:Tsource});
the left-hand side and the first term on the right-hand side are
$\propto \Theta_{\rm s}^3$, whereas the coefficients of the matrix
$\tens{C}\propto \Theta_{\rm s}^4$. This then implies that the accuracy of
the flexion estimates does not depend on the magnitude of the flexion
and the source size individually, but only on the product
$G_n\Theta_{\rm s}$. Therefore, the following results are quoted
always in terms of this product.

\begin{figure}
\centerline{\scalebox{0.5}{\includegraphics[angle=270,scale=1.4]{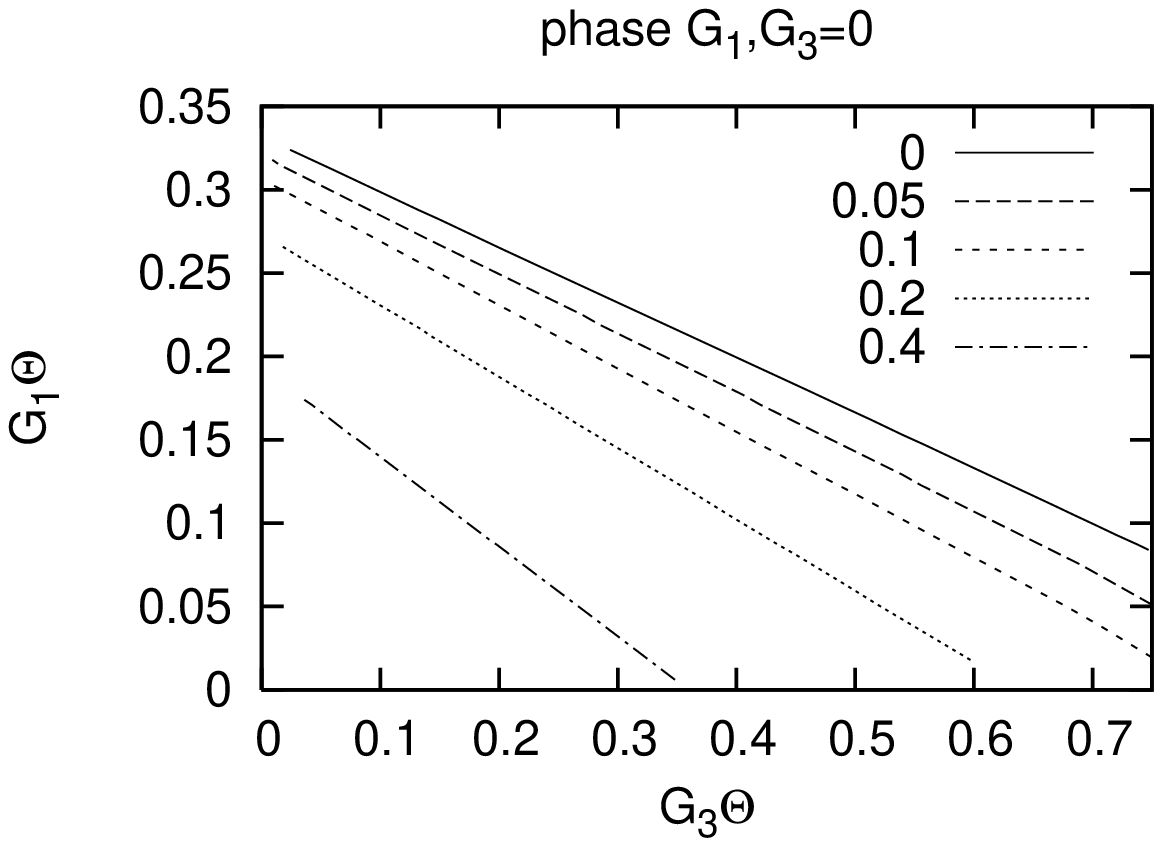}
\vspace*{10pt}\includegraphics[angle=270,scale=1.4]{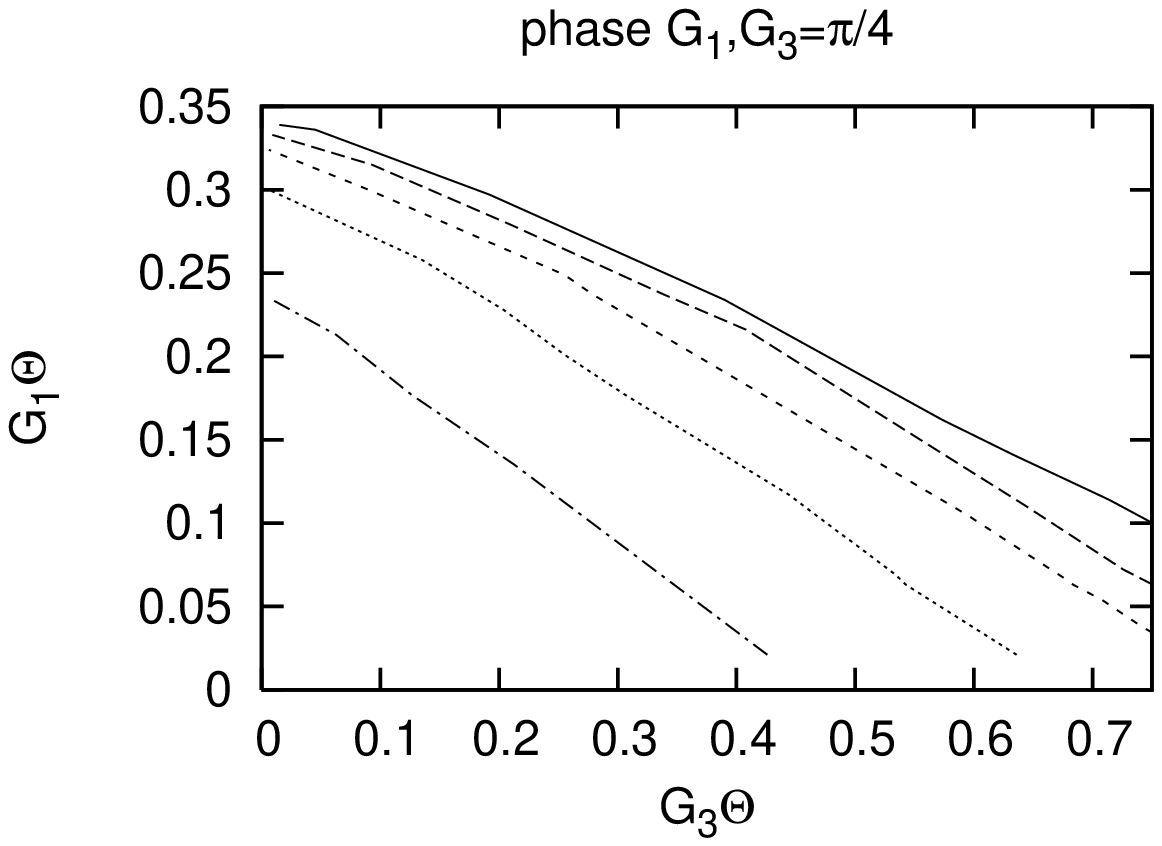}
}}
\centerline{\scalebox{0.5}{\includegraphics[angle=270,scale=1.4]{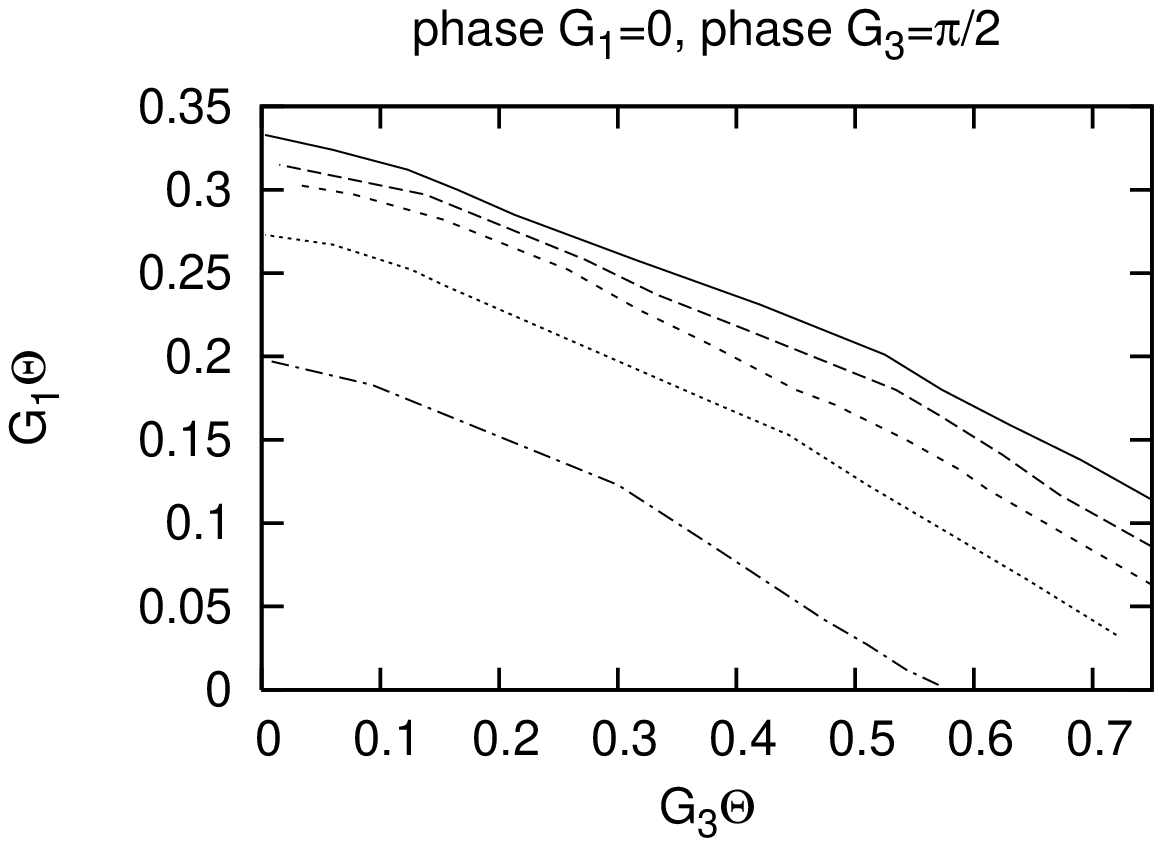}
\vspace*{10pt}\includegraphics[angle=270,scale=1.4]{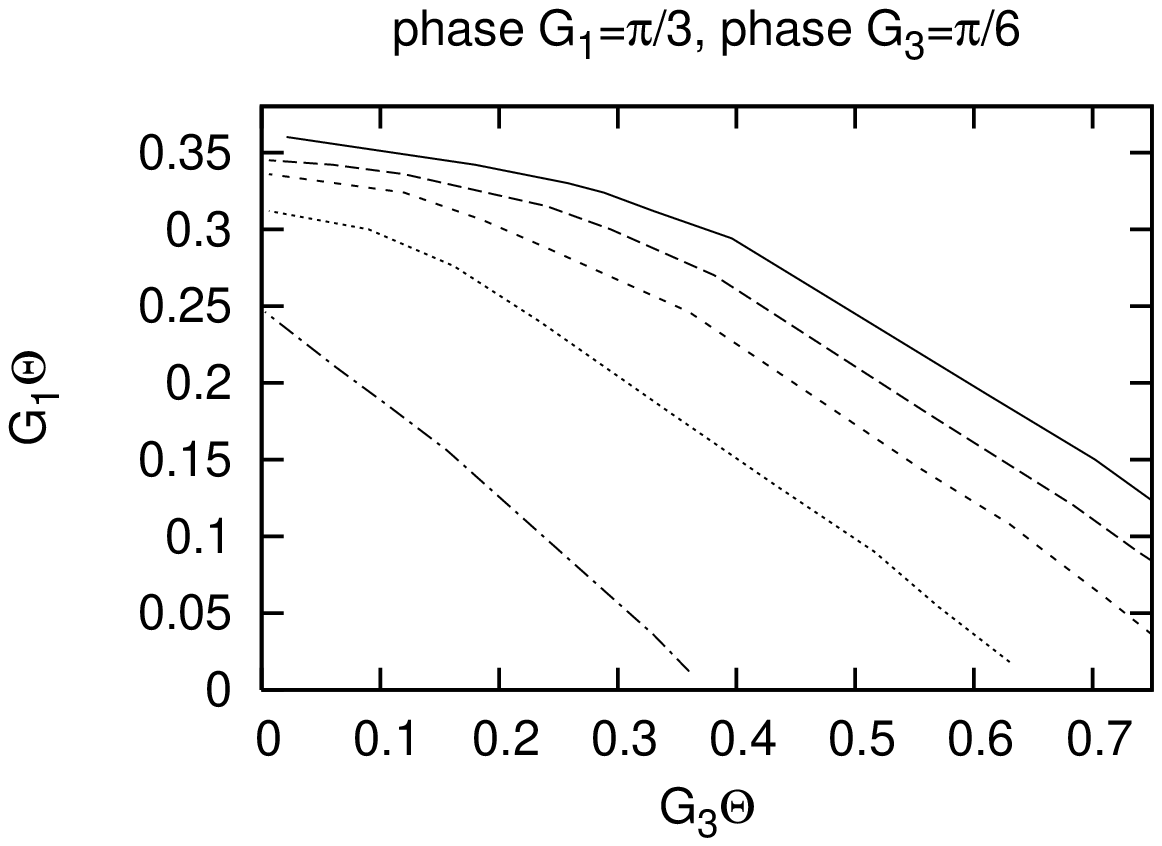}
}}
\caption{
Constrains on the combination of source size and reduced flexion for
the validity of the concept of flexion. Each curve shows the dividing
line between a circular source of limiting isophote $\Theta$
being cut by a caustic (above the curve) or not
(below the curve); in the former case, the assumptions underlying the flexion
concept break down. The different curves in each panel are for
different values of $g$, chosen as $g=0.4,0.2,0.1,0.05,0$, as
indicated by different line types. Without loss of generality, we
choose $g$ to be real and non-negative.
The four panels differ in the phase
of the reduced flexion, as indicated. E.g., in the upper left panel,
the phases of $G_1$, $G_3$ are the same as that of $g$}
\label{fig:cc}
\end{figure}

\subsection{\label{Sc:5.4}Multiple images, and the breakdown of flexion}
As we mentioned before, the lens equation (\ref{eq:lenseq}) can give
rise to multiple images. As can be seen from the example given after
(\ref{eq:lenseq}), if the flexion is sufficiently small, all but one
of these multiple images will be located at a large distance from the
origin, and the central image of an extended source will be
isolated. In this case, this central, or primary, image (the shape of
which we measure here) is not crossed by a critical curve, and thus
the source is not crossed by a caustic. The multiple images at large
distances from the origin then result from the low-order Taylor
expansion of the lens equation, which most likely breaks down at these
image positions anyway; hence, these additional images are of no
relevance.  If, however, the flexion becomes sufficiently large -- or
if the source is large enough -- this is no longer the case, and the
multiple images of an extended source will merge. If that happens, the
whole method of determining shear and derivatives thereof from
brightness moments will break down. This can be most easily seen by
considering the caustic curve cutting the source. Different parts of
the source will be mapped onto a different number of image points in
the lens plane, and the caustic curve introduces a direction into the
situation. Hence, the assumption of an isotropic orientation of
sources can no longer be employed. Mathematically, this can be seen
from (\ref{eq:Momdef}); there, the transformation between source and
image plane no longer is correct if multiple images do occur. More
precisely, the transformation between source and image coordinates in
the calculation of the brightness moments implicitly assumes that
within the limiting isophote of the primary image, the lens equation is
invertible.  Owing to what was said above, the condition that the
central image is isolated (so that locally no multiple images occur)
can be expressed solely by the products $G_n \Theta_{\rm s}$. These
products approximately measure the fractional change of the reduced
shear across the image of a source.

In our simulations we can check whether a critical curve crosses our
central image, just by controlling the sign of the Jacobian
determinant (the true one, not the linear approximation eq.\ts
\ref{eq:detAhat}). If the source size becomes too large, some points in
the image will have a negative Jacobian. In the Appendix B, we
consider the critical curves and caustics of the lens equation
(\ref{eq:lenseqred}), which allows us to determine the regions in
flexion space where no local multiple imaging occurs. Some examples of
this are plotted in Fig.\ts\ref{fig:cc}. Each panel shows the dividing
line between parameter pairs $(G_1\Theta,G_3\Theta)$ for a circular
source of limiting isophotal radius $\Theta$; below the curves, no
local multiple images occur, whereas for parameter pairs above the
lines, the flexion formalism using moments necessarily breaks
down. The different lines in each panel correspond to different values
of $g$. The occurrence of critical curves also is the reason why we
truncated the intrinsic ellipticity distribution of the sources in the
simulations, since in the limit of $|\chi^{\rm s}| \to 1$, keeping the
source area fixed, there will be orientation angles for which the
source will hit a caustic.

\subsection{Estimates of the accuracy}
We now present some results of our numerical simulation regarding the
accuracy with which the reduced shear and flexion can be obtained with
our moment approach. For given input values of $g$, $G_1$ and $G_3$,
we either measure the brightness moments for a single circular source,
or average the results over an ellipticity distribution, as described
above. It should be noted that we have to deal with a 5-dimensional
parameter space, namely the 3 complex parameters $g$, $G_1$ and $G_3$,
minus one overall phase that can be chosen, e.g., to make $g$ real and
positive. Thus, instead of sampling the parameter space comprehensively,
we only give a few selected results.

\begin{figure}
\centerline{\scalebox{0.5}{\includegraphics[angle=0,scale=1.4]{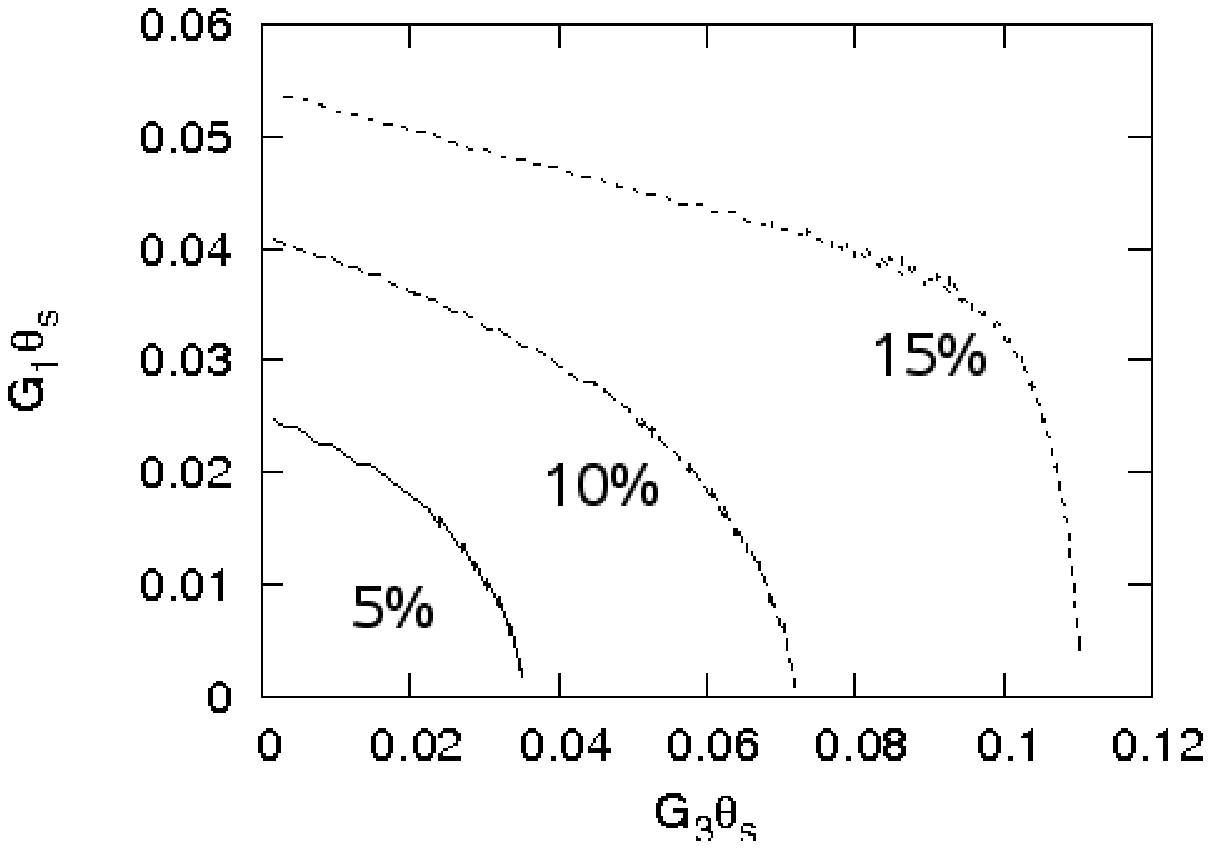}
\hspace*{12pt}\includegraphics[angle=0,scale=1.4]{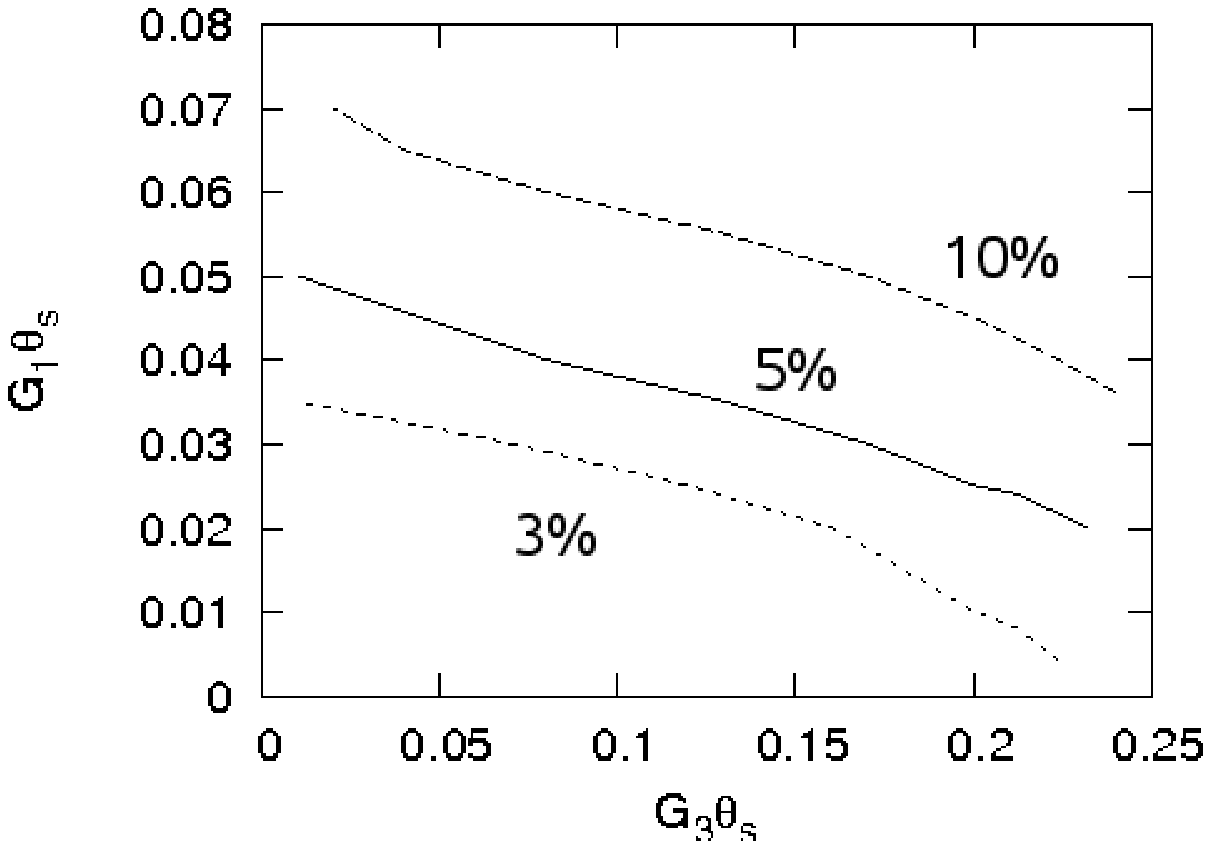}}}
\caption{
Accuracy of the estimates for reduced shear and flexion. The left
panel shows contour of constant fractional error of $5\%$, $10\%$ and
$15\%$, on the estimate of the reduced shear $g$, as a function of
$G_i\Theta_{\rm s}$, where we chose $g=0.05$ as input value, and
assumed the phases of $G_1$, $G_3$ to be the same as that of $g$. The
estimate was obtained by solving the iteration equations given in
Sect.\ts\ref{Sc:5}. The right panel shows the fractional error levels
at 3, 5, and 10\% for the reduced flexion, as quantified by
(\ref{eq:DeltaG}), where the estimate was obtained again with the
iterative procedure. In both cases, we assumed circular sources}
\label{fig:error}
\end{figure}

We start by considering a circular source, and determine the effect of
flexion on the determination of the reduced shear. The left-hand panel
of Fig.\ts\ref{fig:error} shows contours of constant fractional
deviation $\Delta g/g$, in the flexion parameter plane. Here it is
assumed that the phase of both flexion components is the same as that
of $g$ (as would be the case in an axially-symmetric lens
potential). Errors of order 5\% occur already for
$\sqrt{|G_1^2+G_3^2|}\Theta_{\rm s}\sim 0.03$, and the fractional error
increases approximately linearly with the strength of flexion (or with
the source size), although it does not scale equally with both flexion
components.  The reason for this effect has been mentioned before --
flexion affects the transformation between source and image quadrupole
moments, as can be seen in (\ref{eq:Q2s}).

\begin{figure}
\centerline{\scalebox{0.5}{\includegraphics[angle=0,scale=1.4]{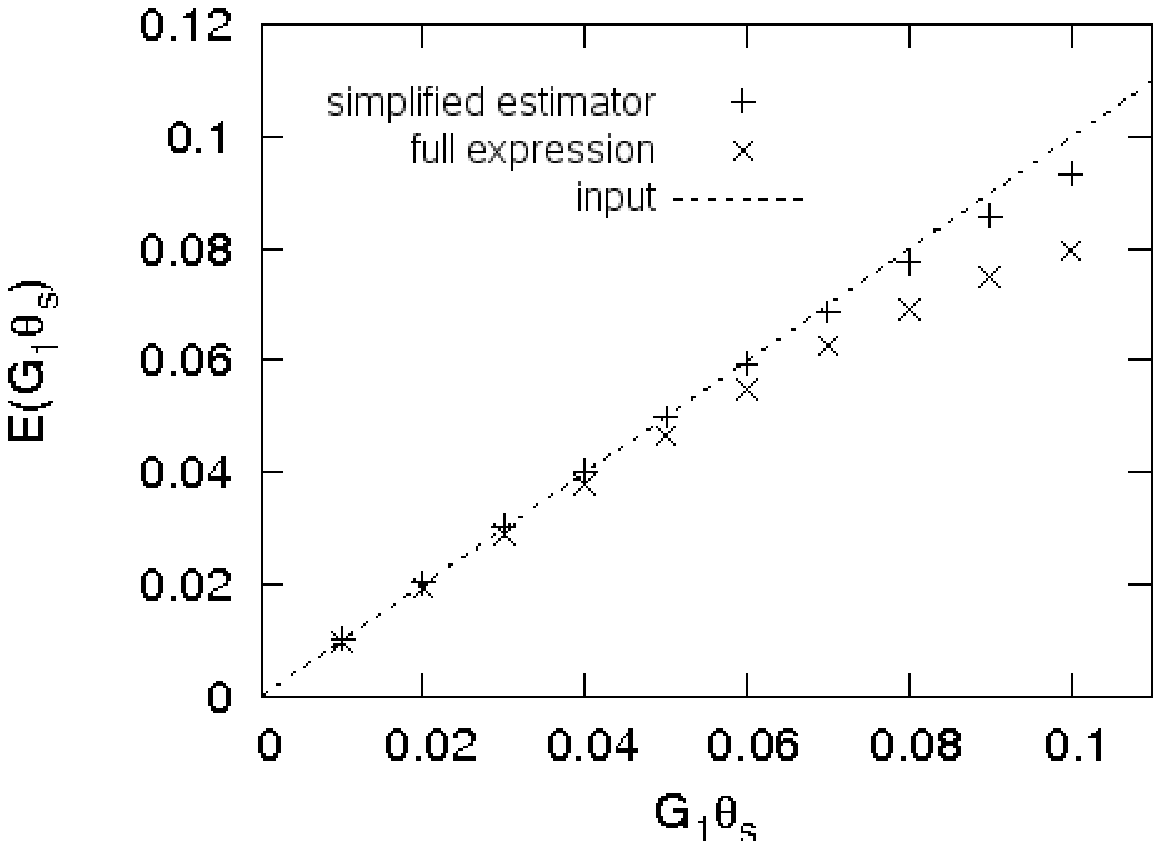}
\hspace*{12pt}\includegraphics[scale=1.4,angle=0]{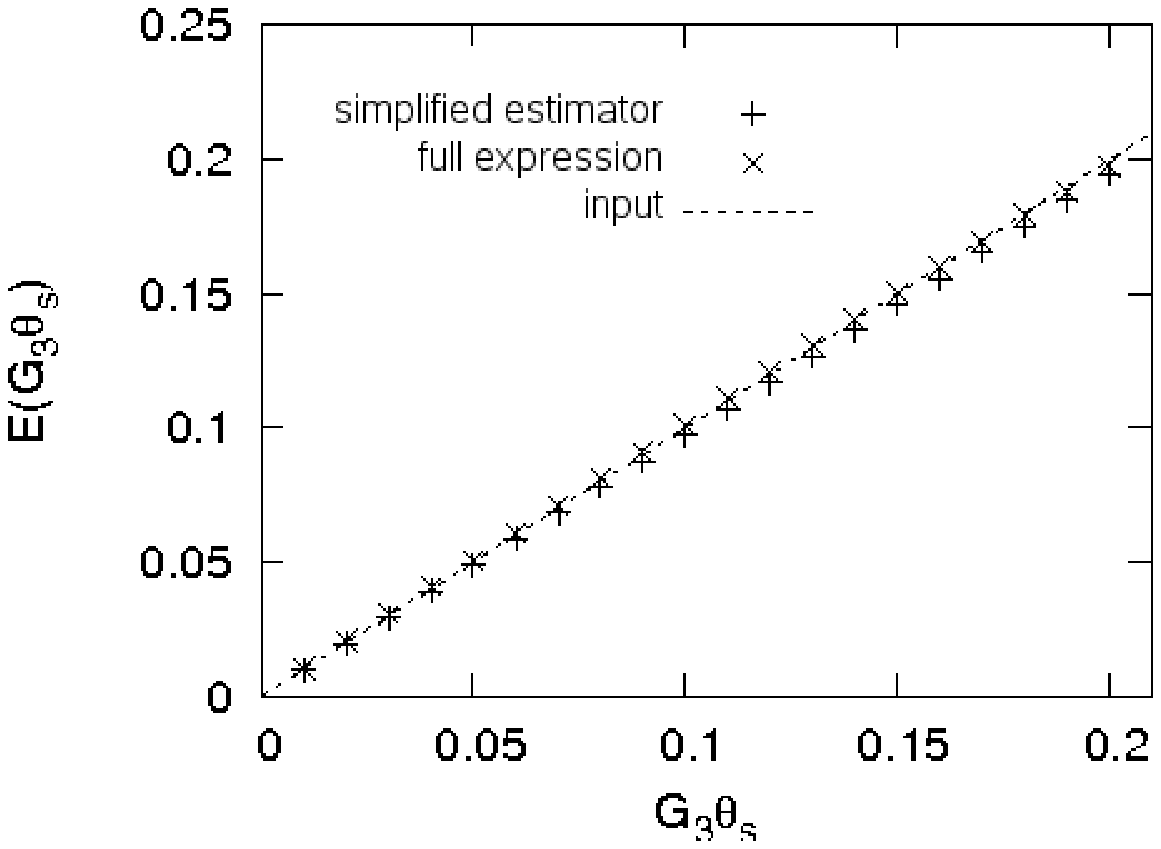}}}
\caption{
Comparison of the reduced flexion estimators (\ref{eq:eg13}) with the
full expression (\ref{eq:g13}) and the input value.  The horizontal
and vertical axis show $G_i \Theta_{\rm s}, i=1,3$. For both
panels, we take $g=0.05$, and $G_3=0$ ($G_1=0$) for the left (right)
panel.  The line indicates the input value, the plus symbols show the
simplified reduced flexion estimate (\ref{eq:eg13}), and the crosses
result from the full expression of reduced flexion
(\ref{eq:g13}). As can be seen from the left-hand panel, the full
estimator for the reduced flexion yields a more biased result that the
approximate expression (\ref{eq:eg13}); we have not found a reasonable
explanation for this behavior}
\label{fig:expG}
\end{figure}

In Fig.\ts\ref{fig:expG}, we show the expectation value of the reduced
flexion components, as a function of the input flexion. The
expectation value has been determined by averaging over an isotropic
ensemble of elliptical sources, as described before. The left and
right panel show the behavior of the expectation value of $G_1$ and
$G_3$, respectively, where the other flexion component was set to
zero. The dashed curve shows the identity, the plus symbols were
obtained by using the approximate estimator (\ref{eq:eg13}), whereas
the crosses show the expectation values as obtained by employing the
full expression (\ref{eq:g13}), where the corresponding value of $g$
was obtained by the iterative process described in
Sect.\ts\ref{Sc:5}. It is reassuring that the expectation value
closely traces the input value, i.e., that the estimates have a fairly
small bias. Furthermore, we see that the approximate estimator
(\ref{eq:eg13}) performs remarkably well. It is seen that the
estimates for $G_3$ behave better than those for $G_1$. This can also
be seen from the right-hand panel of Fig.\ts\ref{fig:error}, where we
plot contours of constant fractional error
\be
\Delta G:=\sqrt{\abs{{\Delta G_1\over G_1}}^2
               +\abs{{\Delta G_3\over G_3}}^2}\;,
\elabel{DeltaG}
\ee
where $\Delta G_n$ is the deviation of the estimate of $G_n$ from its
input value. For simplicity, we have assumed a circular source. We see
that the accuracy decreases much faster with increasing $G_1$ than
with increasing $G_3$. The reason for that may be related to the fact
that the estimator of $G_1$ is more strongly affected by the
non-linearity of the equations, as can also be seen in
(\ref{eq:eg13long}).

\section{\label{Sc:7}Conclusions and further work}

In this paper, we have studied the effect of flexion in weak
gravitational lensing. The main results are summarized as follows:
\begin{itemize}
\item
Owing to the mass-sheet degeneracy, flexion itself cannot be
determined, but only reduced flexion. We have therefore written
the second-order lens equation (which contains the derivatives of the
reduced shear, i.e., flexion) as well as the relations between the
brightness moments of source and image strictly in terms of the
reduced shear and the reduced flexion. 
\item
We pointed out that a general flexion field can be decomposed into a
pair of components which is due to a shear field, i.e., its
derivatives, and a pair of components not related to shear. The former pair
can be further separated into flexion due to an E- and B-mode
shear, with only the E-mode flexion expected to arise from gravitational
lensing. For the second pair of components, no physical interpretation
is available; if they arise in measurements, they are most likely due
to noise or intrinsic shape effects of sources.
General relations to separate these components are given.
\item
We derive the relations between low-order brightness moments of source
and image, taking into account that the presence of flexion leads to
a centroid shift, and it also affects the relation between second-order
brightness moments -- and thus the estimate of the reduced
shear. Hence, the presence of flexion has an impact on the shear
measurements. Starting from these moment equations, we obtain
approximate estimates for the reduced shear and flexion.
\item
We point out a limit where the flexion formalism ceases to be valid,
namely when the product of source size and flexion is sufficiently
large that parts of the source are multiply imaged locally, i.e.,
where a caustic cuts through the source. We have quantified this with
numerical simulations, and also have given a complete classification
of the critical curves of the second-order lens equation employed in
flexion studies.
\item
We have performed a number of numerical experiments to study the bias
of the reduced shear and flexion estimators. However, due to the high
dimensionality of parameter space, no comprehensive study has been
presented here. We also point out that only the product of flexion and
source size matters in the accuracy of estimates.
\end{itemize}
The possible occurrence of critical curves in highly distorted images
may provide a serious obstacle to applications of flexion. Perhaps the
most promising application of flexion measurements are those in
regions where the shear field varies on small scales, i.e., close to
galaxies (and thus can be used for galaxy-galaxy lensing) or in the
inner regions of clusters. However, if one finds a strongly distorted
image of a background galaxy as in the case of the arclet A5 in Abell
370 (Fort et al.\ 1988), how can one be sure that it is not due to a
merged double image of the source? Using flexion for studying
small-scale structure in mass distributions can therefore be affected
by the occurrence of multiple imaging.

Similar to the situation in shear measurements, the moment approach
for flexion as presented here must be modified in several ways to be
applicable to real data. First, brightness moments must be weighted in
order not to be dominated by the very noisy outer regions of the
image. As is known from shear measurements, such a weighting affects
the relation between source and image brightness moments. Secondly,
one needs to account for the effects of a point-spread function. Both
of these modifications have been successfully achieved for
second-order brightness moments by Kaiser et al.\ (1995; see also
Luppino \& Kaiser 1997). Goldberg \& Leonard (2007) consider these
effects in the context of flexion. It should be noted, though, that
their consideration of the PSF effects is restricted to unweighted
moments, for which these effects are given by a simple convolution. In
the case of weighted brightness moments, however, the PSF effects are
much more subtle, and we expect that a formalism in analogy to Kaiser
et al.\ (1995) needs to be developed -- and that this formalism for
higher-order brightness moments will be considerably more difficult
than for the shear case.  

But even disregarding these complications, the present paper only
scratches the surface in investigating estimators for reduced flexion
and their properties. As mentioned before, the second-order lens
equation contains five essential parameters. The bias of an estimator
for reduced shear and flexion will depend on these parameters, as well
as on the intrinsic ellipticity (and higher-order moments) distribution
of sources. One might ask whether it is possible to find an unbiased
flexion estimator, such as was possible to construct for the reduced
shear.  Unfortunately, we have been unable to make analytic progress:
even for a circular Gaussian source, the brightness moments of the
image cannot be calculated analytically. Our ray-tracing algorithm
with which we conducted our numerical simulations is almost certainly
sub-optimal; a more advanced method should be developed to reduce the
numerical efforts in calculating brightness moments. Beside the bias,
it would be interesting to calculate the variance of the various
estimators, or more precisely, their covariance.

It may turn out that measurements of flexion, and PSF
corrections, are more conveniently done with shapelets, as was
originally considered by Goldberg \& Bacon (2005), Bacon et al.\
(2006) and Massey et al.\ (2006). Even if this turns out to be the
case (see Leonard et al.\ 2007 for an application of flexion
measurements in the galaxy cluster A\ts 1689), the moment approach
provides a more intuitive picture of the effects of flexion. In
addition, the weak lensing community has profited substantially from
the existence of several different methods to measure shear (see
Heymans et al.\ 2006; Massey et al.\ 2007 for the first results of a
comprehensive Shear TEsting Programme, in which these various  methods
are studied and compared); therefore, the development of different
techniques for measuring flexion will certainly be of interest once
the flexion method will be put to extensive use.

\section*{Acknowledgments}
We thank Jan Hartlap and Ismael Tereno for useful comments on this
paper. This work was supported by the Deutche Forschungsgemeinschaft
under the project SCHN 342/6--1 and the TR33 `The Dark Universe'.
XE was supported for this research through a stipend from the
International Max-Planck Research School (IMPRS) for Radio and
Infrared Astronomy at the University of Bonn.

\begin{appendix}
\section{\label{Sc:A}The matrix $C$}

In this Appendix, we list the coefficients of the matrix $\tens{C}$
which occurs in (\ref{eq:Tsource}):
\bea
4(1-gg^*) C_{11} &=& -2g F_2^*
+(9g g^*-3) F_0 +6 g^*(1-2 g g^*) F_2+g^{*2}(5 g g^*-3) F_4\nonumber\\
&+&6 g Q_2^* Q_0
-12 g g^* Q_0^2 +(3 -9 g g^*) Q_2^* Q_2
+ 6 g^* (4 g g^*-1) Q_0 Q_2+3 g^{*2}(1- 3 g g^*)  Q_2^2   \nonumber\\
4(1-gg^*) C_{12} &=& 5 g  F_4^* -2(5+6 g g^*) F_2^*
+9 g^*(3+g g^*)  F_0 -2 g^{*2}(12+g g^*) F_2 +7 g^{*3} F_4
-9 g Q_2^{*2} +6(3 +4 g g^*) Q_2^* Q_0 \nonumber\\
&-&12 g^*(3+g g^*) Q_0^2-3 g^*(5+3 g g^*)  Q_2^* Q_2
+6 g^{*2}(8+g g^*) Q_0 Q_2-15 g^{*3}  Q_2^2   \nonumber\\
4(1-gg^*) C_{13} &=& -7 F_4^* +26 g^* F_2^*
-36 g^{*2}  F_0 +22 g^{*3} F_2 -5 g^{*4} F_4\nonumber\\
 &+& 15 Q_2^{*2} -54 g^*  Q_2^* Q_0
 +48 g^{*2}  Q_0^2 +24 g^{*2}  Q_2^* Q_2
 -42 g^{*3}  Q_0 Q_2 +9 g^{*4}  Q_2^2   \nonumber\\
4(1-gg^*) C_{14} &=& -2 g^* F_4^* +6 g^{*2} F_2^*
- 6 g^{*3} F_0 +2 g^{*4}  F_2  \nonumber\\
&+&6 g^*  Q_2^{*2} -18 g^{*2}  Q_2^* Q_0
+12 g^{*3}  Q_0^2 +6 g^{*3}  Q_2^* Q_2
 -6 g^{*4} Q_0 Q_2     \nonumber\\
4(1-gg^*) C_{21} &=&  2 g^2   F_2^*
-6 g^2 g^* F_0+[4 g g^*(1+g g^*)-2] F_2+2 g^*(1-2 g g^*) F_4\nonumber\\
&-&6 g^2   Q_2^* Q_0
 +4 g (1+2 g g^*) Q_0^2 +6 g^2 g^*  Q_2^* Q_2
+[2-4 g g^*(3+2 g g^*)]  Q_0 Q_2+2 g^* (4 g g^*-1) Q_2^2 \nonumber\\
4(1-gg^*) C_{22} &=& -5 g^2  F_4^* +2 g (7+4 g g^*) F_2^*
-3[3+g g^*(8+g g^*)] F_0+2 g^*(8+5 g g^*) F_2-7 g^{*2}  F_4
+ 9 g^2  Q_2^{*2}
\nonumber\\
-2 g (13+8 g g^*)\!\!\!\!\!\!\!\!\!\!&& Q_2^* Q_0 +4[3+g g^*(8+g g^*)] Q_0^2
+[5+g g^*(16+3 g g^*)] Q_2^* Q_2
-2 g^*(16+11 g g^*) Q_0 Q_2 +15 g^{*2} Q_2^2   \nonumber\\
4(1-gg^*) C_{23} &=& 7g F_4^* -2(4 +9 g g^*)  F_2^*
+3 g^*(7+5 g g^*) F_0 -2 g^{*2}(9+2 g g^*) F_2+5 g^{*3} F_4
-15 g Q_2^{*2}
\nonumber\\
&+&(16+38 g g^*) Q_2^* Q_0 - 4 g^*(7 +5 g g^*) Q_0^2
-  g^*(13+11 g g^*) Q_2^* Q_2
+2 g^{*2}(17+4 g g^*) Q_0 Q_2-9 g^{*3} Q_2^2   \nonumber\\
4(1-gg^*) C_{24} &=& (3 g g^*-1) F_4^*-6 g g^{*2} F_2^*
+3 g^{*2}(1 + g g^*) F_0 -2 g^{*3} F_2  \nonumber\\
&+&(1 -7 g g^*) Q_2^{*2}+2 g^*(2+7 g g^*) Q_2^* Q_0
- 4 g^{*2}(2+g g^*)  Q_0^2-3 g^{*2}(1+g g^*) Q_2^* Q_2
+6 g^{*3} Q_0 Q_2   \nonumber
\eea
The other eight elements follow trivially from the foregoing ones,
since the second half of the matrix is just the complex conjugate one
of the first half, i.e., $C_{44}=C_{11}^*$, $C_{34}=C_{21}^*$ etc.,
or in general, $C_{ij}=C_{5-i,5-j}^*$.

\begin{figure}
\centerline{\scalebox{1}{\includegraphics[angle=0,scale=0.8]{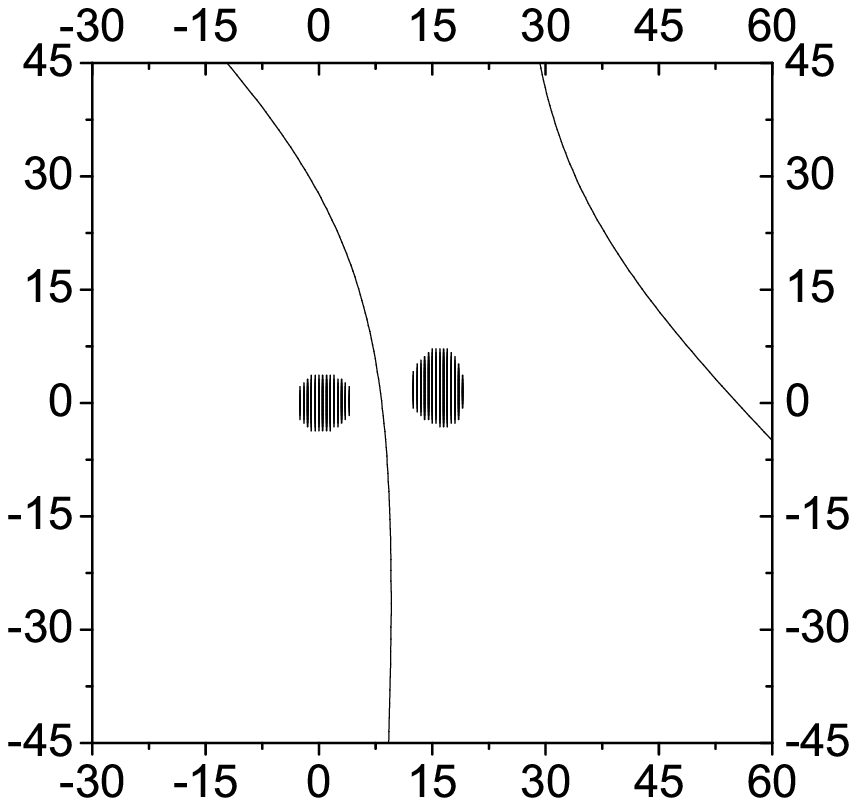}
\vspace*{10pt}\includegraphics[angle=0,scale=0.8]{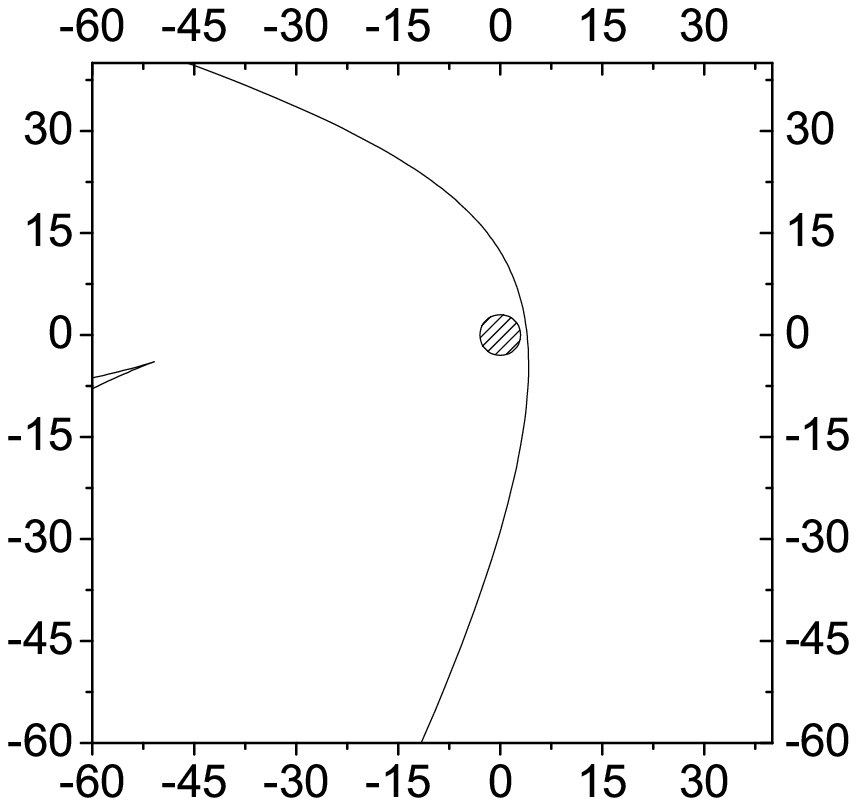}}
}
\caption{The critical curves (left-hand panel) and caustics
(right-hand panel) of the lens equation (\ref{eq:lenseqred}) for the
cases of hyperbolic critical curves, as described in
Sect.\ts\ref{sc:B2}. The parameters chosen here are $g=0.05$,
$G_1=0.07+0.015{\rm i}$, $G_3=0.03+0.005{\rm i}$. A circular source 
is mapped onto two images, as indicated. If the
source size were increased, it would hit the caustic, the two images
would merge, and the flexion concept would break down. The unit of
the reduced flexion is the inverse of the unit in which coordinates
are measured
}
\label{fig:hyper}
\end{figure}

\begin{figure}
\centerline{\scalebox{1}{\includegraphics[angle=0,scale=0.8]{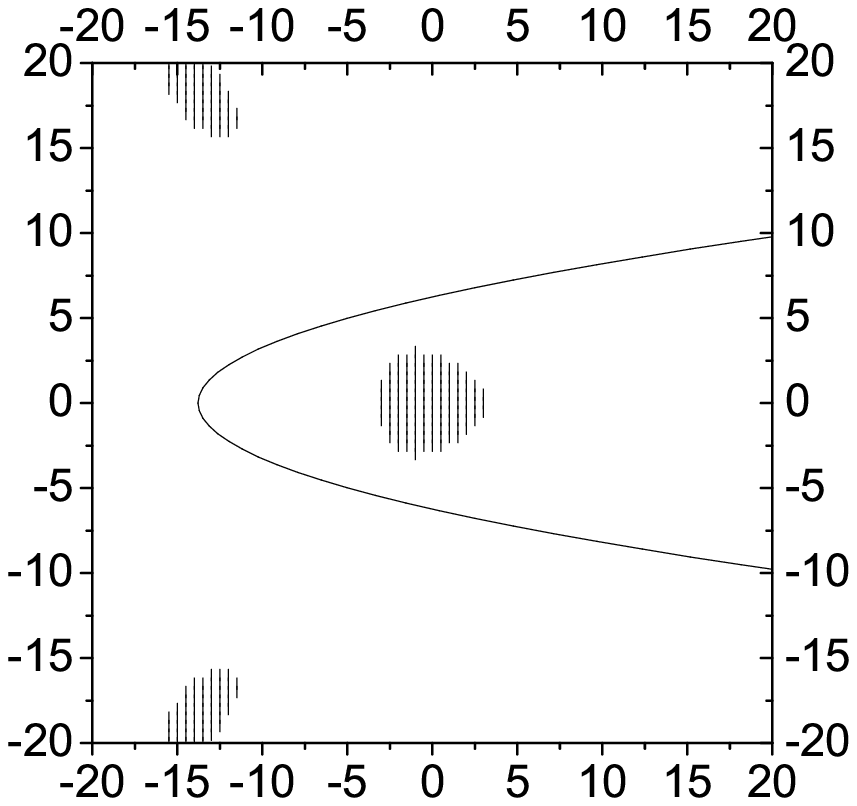}
\vspace*{10pt}\includegraphics[angle=0,scale=0.8]{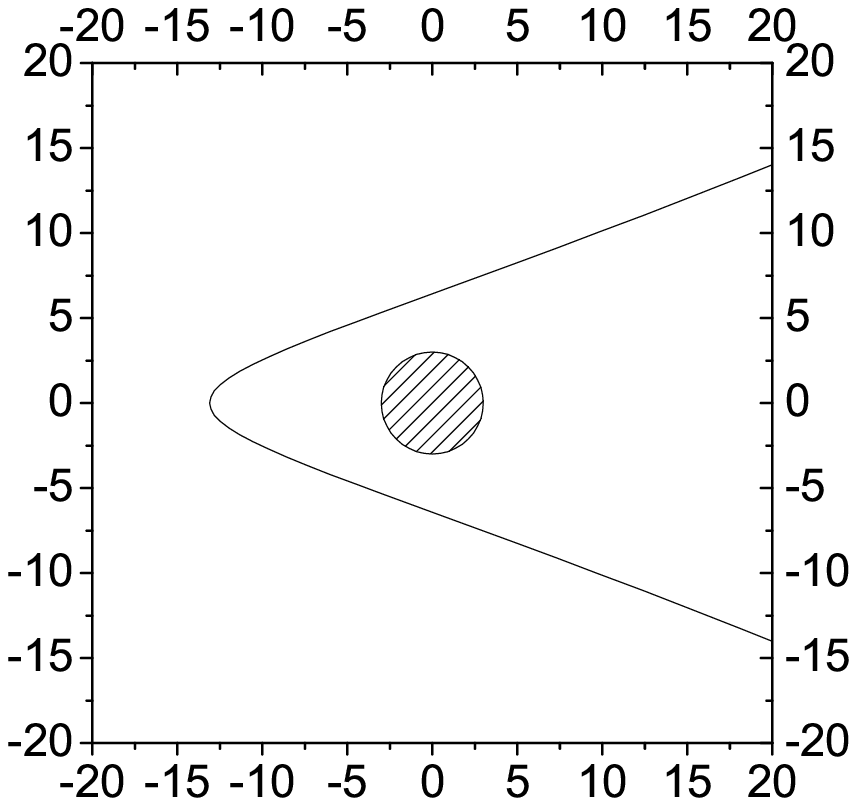}}
}
\caption{Same as Fig.\ts\ref{fig:hyper}, but for the parabolic case,
with parameters
$g=0.05$, $G_1=-0.04$, $G_3=0.112$
}
\label{fig:para}
\end{figure}

\begin{figure}
\centerline{{\includegraphics[angle=0,scale=0.8]{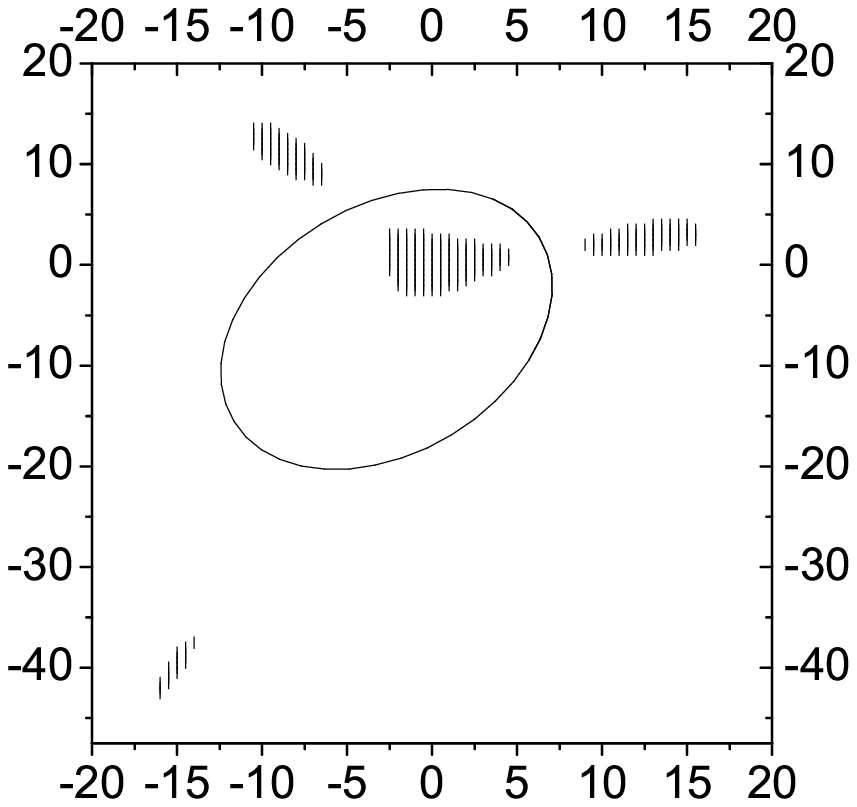}
\vspace*{10pt}\includegraphics[angle=0,scale=0.8]{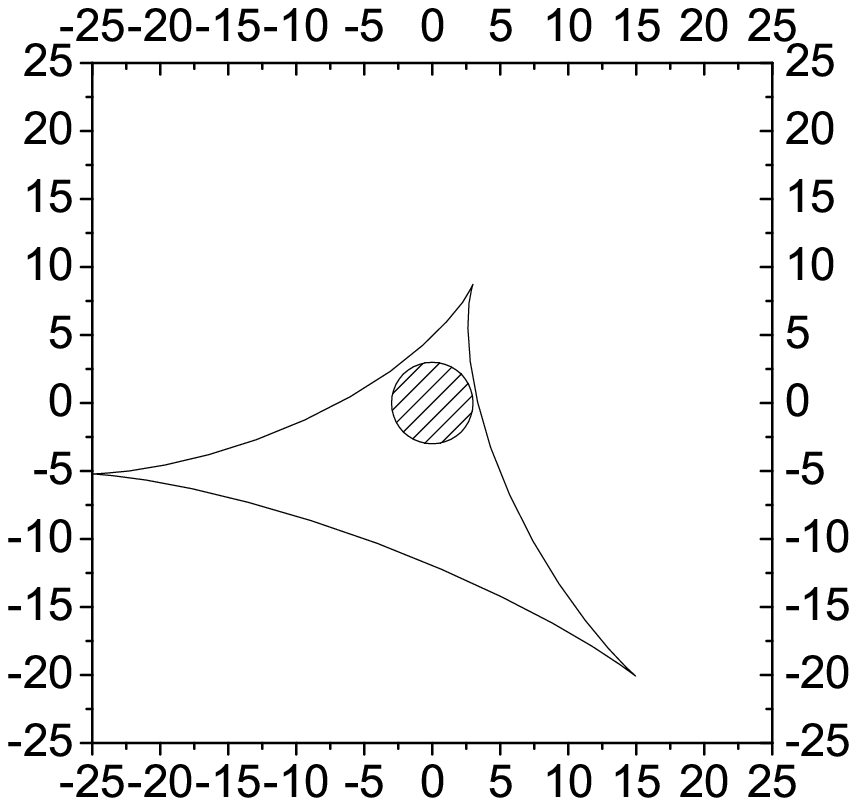}}
}
\caption{Same as Fig.\ts\ref{fig:hyper}, but for the elliptical case,
with parameters
$g=0.05,G_1=0.015+0.035{\rm i}, G_3=0.19+0.105{\rm i}$
}
\label{fig:ellipse}
\end{figure}

\section{\label{Sc:B}Critical curves and caustics}
In this Appendix we consider the critical curves of the lens equation
(\ref{eq:lenseqred}). For this, we need to derive the full Jacobian,
which can most easily be obtained from considering $\theta$ and
$\theta^*$ as independent variables, and then use
$\partial/\partial\theta_1 = \partial/\partial\theta +
\partial/\partial\theta^*$, $\partial/\partial\theta_2 = 
{\rm i}\rund{\partial/\partial\theta -
\partial/\partial\theta^*}$, which can be inverted to yield 
$\partial/\partial\theta=\nabla_{\rm c}^*/2$, 
$\partial/\partial\theta^*=\nabla_{\rm c}/2$. With these relations, one
finds that $\det\A=(\partial\beta/\partial\theta)
(\partial\beta^*/\partial\theta^*) -
(\partial\beta/\partial\theta^*)
(\partial\beta^*/\partial\theta)
=
\rund{\nabla_{\rm c}^*\beta\,\nabla_{\rm c}\beta^*
-\nabla_{\rm c}\beta\,\nabla_{\rm c}^*\beta^*}/4$. Carrying out these
derivatives, the Jacobian becomes
\be
\det\A=1-g g^* -\eta^*\theta-\eta\theta^* +
A^*\theta^2+B\theta\theta^*
+A(\theta^*)^2\;,
\ee
with
\be
A=4\rund{\Psi_1^2-\Psi_1^*\Psi_3}\; ;\quad
B=4\rund{\Psi_1\Psi_1^*-\Psi_3\Psi_3^*} \; ;\quad
\eta=4\Psi_1+2g\Psi_1^*+2 g^*\Psi_3\;.
\ee
Note that $A$ is a spin-2 quantity, whereas $B$ is a real scalar,
i.e., has spin-0.
In the generic case, the critical curves ($\det\A=0$) are conical
sections, which may be degenerate, though. We will now perform a
complete classification of cases that can occur, as well as to derive
the critical curve(s) in parametric form. As we shall see, the type of conical
section is determined, amongst other parameters, by the discriminant
\be
\Delta=B^2-4 A A^*\;.
\ee

\subsection{Zero discriminant}
We start with the case that $\Delta=0$, which implies $B^2=4 A A^*$,
or $B=\pm 2|A|$. The case $A=0=B$ either implies that
$\Psi_1=0=\Psi_3$, in which case also $\eta=0$ so that no critical
curves occur, or that $\Psi_3=\Psi_1^2/\Psi_1^*$, for which $\eta\ne
0$ in general. In this latter case, the critical curve is a straight
line, satisfying $\eta^*\theta+\eta\theta^*=1-g g^*$. As can be seen
by inspection, it reads
\be
\theta={1-g g^*\over 2\eta^*}+{\rm i}\lambda\eta\;,\quad
-\infty<\lambda<\infty\;.
\ee
If $A\ne 0$, the phase of $A$ is defined. Since it is a spin-2
quantity, we write $A=|A|\,{\rm e}^{2{\rm i}\vp_A}$. Furthermore, we
introduce the rotation $\theta=x\,{\rm e}^{{\rm i}\vp_A}$. Then the
equation for the critical curve reads
\be
\rund{x\pm x^*}^2=\nu^* x+\nu x^*+{g g^* -1\over |A|}\;,\quad
{\rm with}\; \nu={\eta\,{\rm e}^{-{\rm i}\vp_A} \over |A|} \;,
\ee
and the sign on the left-hand side of the equation depends on the sign
of $B$, where we used $B=\pm 2|A|$. The parametric form of the
critical curve, which takes the form of a parabola, can then be
written as
\be
\theta={2\,{\rm e}^{{\rm i}\vp_A}\over (\nu^*-\nu)}
\rund{2\lambda^2-\lambda \nu+{1-g g^*\over 2|A|}} \; ; \quad
\theta={2\,{\rm e}^{{\rm i}\vp_A}\over (\nu^*+\nu)}
\rund{{1-g g^*\over 2|A|}-{\rm i}\lambda\nu - 2\lambda^2}\;,
\ee
where the first (second) equation applies for $B>0$ ($B<0$). Note that
the parabola degenerates into a straight line if $\nu$ is real (for
$B>0$) or purely imaginary (for $B<0$).

\subsection{\label{sc:B2}Non-zero discriminant}
If $\Delta\ne 0$, we can perform a translation to eliminate the linear
term in $\det\A$. Hence we define $\theta=\theta_0+\vt$ and choose
$\theta_0$ such that terms linear in $\vt$ vanish. We then obtain 
for $\theta_0$ and for the critical curve condition
\be
\theta_0={B\eta-2 A\eta^*\over \Delta}\;;\quad
A^* \vt^2+B\vt\vt^*+A(\vt^*)^2=C\;,
\elabel{theta-0}
\ee
with
\be
C={B \eta\eta^*-A(\eta^*)^2-A^*\eta^2\over \Delta}+g g^*-1
=-{1\over \Delta}\,\rund{g A^* +g^* A+B}^2 =: -{1\over \Delta}\,V^2\;,
\elabel{Cequation}
\ee
where the second step was obtained by inserting the expression for
$\eta$ in terms of the $\Psi$'s, and in the final one we defined $V$
as the expression in the parenthesis.

As the first case, we consider $A=0$ and $B\ne 0$ (the case $A=0=B$
was treated above), which implies that $\Psi_1=0$ and
$B=-4\Psi_3\Psi_3^*<0$. The equation for the critical curve then
reduces to $B|\vt|^2=C$. Furthermore, $\Delta=B^2$, and $C=-1$. Thus,
the critical curve is a circle of radius $1/(2|\Psi_3|)$ and center
$\theta_0$, or $\theta=\theta_0+{\rm e}^{{\rm i}\lambda}/(2|\Psi_3|)$,
$0\le \lambda<2\pi$. 

We now consider the case $A\ne 0$; then the phase $\vp_A$ of $A$ is defined,
as used before. Introducing a rotation by defining $\vt=x\,{\rm
e}^{{\rm i}{\vp_A}}$, the equation for the critical curve becomes
\be
|A|\eck{x^2+\rund{x^*}^2}+B x x^*
=\rund{B+2|A|}x_1^2+\rund{B-2|A|}x_2^2=C\;.
\elabel{CC2}
\ee
The presence and topology of critical curves now depends on the signs
of $\Delta$ and $C$. We first consider the case $C=0$; then, if
$\Delta>0$, no critical curves occur, except for the isolated point
$x=0$. If $\Delta<0$, the critical curves are two straight lines, as
can be obtained from (\ref{eq:theta-0}): inserting the ansatz
$\vt=\lambda\,{\rm e}^{{\rm i}\zeta}$, one obtains ${\rm e}^{2{\rm
i}(\zeta-\vp_A)} =(-B\pm{\rm i}\sqrt{-\Delta})/(2|A|)$. Thus, the
critical curves are parametrized as
\be
\theta=\theta_0+\lambda\,{\rm e}^{{\rm i}\vp_A}
\sqrt{-B\pm{\rm i}\sqrt{-\Delta}\over 2|A|} \;;\quad
-\infty<\lambda<\infty\;.
\ee
For the case of $C\ne 0$, the consideration of (\ref{eq:CC2}) yields
the result that for $\Delta<0$, the critical curves consist of two
hyperbolae. From (\ref{eq:Cequation}) we see that negative $\Delta$
implies $C>0$. Also note that $\Delta<0$ implies that $2|A|-B>0$,
$2|A|+B>0$. The critical curves then read
\be
\theta=\theta_0+{ {\rm e}^{{\rm i}\vp_A}\, V\over\sqrt{-\Delta}}
\rund{ \pm {\cosh\lambda\over \sqrt{2|A|+B}} +{\rm i}\,
{\sinh\lambda\over \sqrt{2|A|-B}}}\; ;\quad
-\infty<\lambda<\infty \;.
\ee
For the other case, $\Delta>0$, we find from (\ref{eq:Cequation}) that
$C<0$. If $B\pm 2|A|>0$, we then see from (\ref{eq:CC2}) that no
critical curves exist. If $B\pm2|A|<0$, which in particular implies
$B<0$, the critical curve is an ellipse parametrized as
\be
\theta=\theta_0+{ {\rm e}^{{\rm i}\vp_A}\, V\over\sqrt{\Delta}}
\rund{{\cos\lambda\over \sqrt{-2|A|-B}} +{\rm i}\,
{\sin\lambda\over \sqrt{2|A|-B}}}\; ;\quad
0\le \lambda < 2\pi\;.
\ee
This concludes the classification of critical curves of the lens
equation (\ref{eq:lenseqred}). The caustics are obtained by inserting
the parametrized form of the critical curves into the lens
equation. In order to see whether a critical curves cuts through the
primary image of a circular source of outer isophotal radius $\Theta$,
we calculate the minimum value $\beta_{\rm min}$ of $|\beta(\lambda)|$
along the caustics. If $\beta_{\rm min}>\Theta$, the image is not cut
by a critical curve. For an elliptical critical curve, the maximum
source size allowed is $\beta_{\rm min}$; these values are plotted in
Fig.\ts\ref{fig:cc}. In the cases where two critical curves exist
(e.g., two straight lines or hyperbolae), the situation is
slightly more complicated. Consider, e.g., the case of two straight
critical curves. Only those sections of them that are closer to the
origin are relevant for this consideration, since if the primary image
of the source is not cut by these closer sections of critical curves,
it will still be an isolated image; the caustics coming from the outer
sections of the critical curves correspond to multiply imaged source
sections of secondary images. Accounting for this complication, the
maximum sources size have been obtained, as plotted in Fig.\ts\ref{fig:cc}.

\end{appendix}


\end{document}